\documentclass[twocolumn, superscriptaddress, pre, aps, floatfix, longbibliography ]{revtex4-1}

\usepackage{amssymb}
\usepackage{amsmath}
\usepackage{graphicx}
\usepackage{color}
\usepackage{hyperref}

\begin{document}

\author{Marco Antonio Rodr\'iguez Flores}
\email{mj.rodriguezflores@edu.cut.ac.cy}
\affiliation{Department of Electrical Engineering, Computer Engineering and Informatics, Cyprus University of Technology, 33 Saripolou Street, 3036 Limassol, Cyprus}
\affiliation{Social Computing Research Centre (SCRC), Cyprus University of Technology, Limassol, Cyprus}
\author{Fragkiskos Papadopoulos}
\email{f.papadopoulos@cut.ac.cy}
\affiliation{Department of Electrical Engineering, Computer Engineering and Informatics, Cyprus University of Technology, 33 Saripolou Street, 3036 Limassol, Cyprus}
\affiliation{Social Computing Research Centre (SCRC), Cyprus University of Technology, Limassol, Cyprus}

\title{Similarity forces and recurrent components in human face-to-face interaction networks}

\begin{abstract}
We show that the social dynamics responsible for the formation of connected components that appear recurrently in face-to-face interaction networks, find a natural explanation in the assumption that the agents of the temporal network reside in a hidden similarity space. Distances between the agents in this space act as similarity forces directing their motion towards other agents in the physical space and determining the duration of their interactions. By contrast, if such forces are ignored in the motion of the agents recurrent components do not form, although other main properties of such networks can still be reproduced.
\end{abstract}

\maketitle

Understanding the mechanisms that drive the dynamics of face-to-face interaction networks is crucial for better analyses of spreading phenomena. In particular, phenomena that evolve as fast as real-time face-to-face interactions, such as respiratory transmitted diseases, word-of-mouth information transfer and viruses in mobile networks~\cite{BarratApps2015, Masuda2016, HolmeSurvey2015}. Furthermore, deriving efficient epidemic control strategies requires an accurate description of fast-evolving contagions~\cite{BarratApps2015, HolmeStructure2016, HolmeVacc2016, KarsaiContainment, OguraContainment}. However, a complete understanding of the processes responsible for the structural and dynamical properties of face-to-face interaction networks has been an elusive task~\cite{SuneRecurrent, HolmeSurvey2015, BarbosaMobility2017}.

Face-to-face interaction networks portray social interactions in closed settings such as schools, hospitals, offices, etc. A typical representation consists of a series of network snapshots. Each snapshot corresponds to an observation interval, which can span from a few seconds to several minutes depending on the devices used to collect the data~\cite{SocioPatterns, StarniniDevices2017}. The agents (nodes) in each snapshot are individuals and an edge between any two agents represents a direct face-to-face interaction. 

Analyses of such networks have uncovered universal properties, such as the heavy-tailed distributions of the interaction duration and time between consecutive interactions,  cf.~\cite{ConferenceData}. Previous results point to the idea of social attractiveness as a mechanism responsible for these universal properties and for other structural characteristics of the time-aggregated network of contacts, like its degree, weight and strength distributions~\cite{Starnini2013, Starnini2016, StarniniDevices2017}. Specifically, in the attractiveness model~\cite{Starnini2013, Starnini2016} agents have an activation probability $r_i$ and a \emph{global attractiveness} value $a_i$ that are sampled uniformly at random from $[0,1]$. Time is slotted and in each slot each non-interacting agent $i$ is active with probability $r_i$. Active agents perform \emph{random walks} in a closed Euclidean space moving towards a random direction every slot with a constant velocity (displacement) $v$. Agents stop moving to interact whenever they encounter another agent within a threshold distance $d$. The activation probability represents the activeness of each agent in the social event. The global attractiveness of the agents defines an \emph{escaping probability} from the interactions. For instance, an agent $i$ that has stopped moving in order to interact with other agents within distance $d$, can resume mobility with probability $1-\max_{j \in \mathcal{N}_i}\{a_j\}$, where $\mathcal{N}_i$ is the set of agents interacting with $i$~\cite{Starnini2013}. Therefore, longer interactions occur when an individual with a high global attractiveness $a_j$ is involved. 

\begin{figure*}[t]
\centering
\includegraphics[width=18cm]{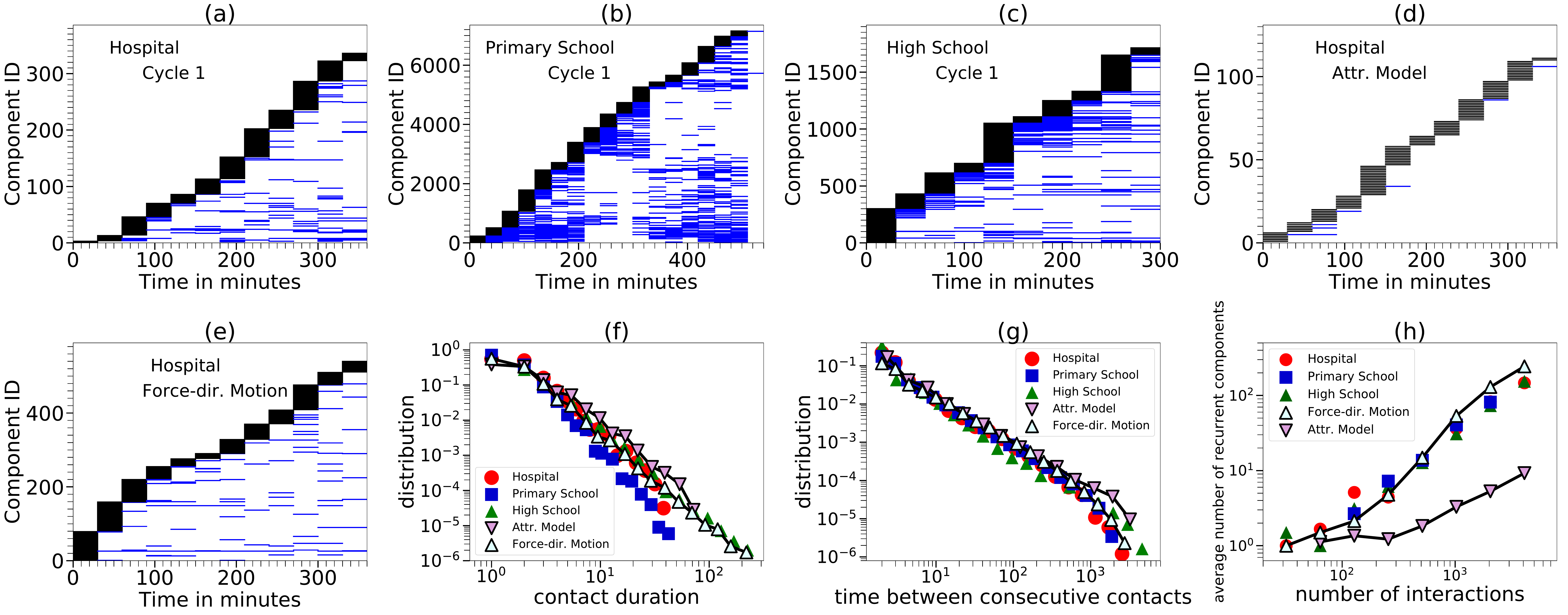}
\caption{Recurrent component patterns and distributions of contact durations and of times between consecutive contacts in three real-world datasets and in simulated networks. 
\textbf{(a-c)} Components found in the first activity cycle of the Hospital, Primary School and High School (6, 8.6 and 5~hours, respectively). 
\textbf{(d)} Components found in a simulation of the attractiveness model with the same duration as in (a). 
\textbf{(e)} Same as (d) but with the FDM (Force-dir.~Motion) model. 
\textbf{(f,~g)}~Distribution of contact duration and of time between consecutive contacts in real and simulated networks. 
\textbf{(h)} Average number of recurrent components where an agent participates as a function of its total number of interactions in real and simulated networks.
The blue lines in (a-e) correspond to \emph{recurrent} components while the black lines to components appearing for the first time, i.e., to the \emph{unique} components. The $x$-axis is binned into 30 minute intervals, while the $y$-axis shows the component IDs observed in each bin; all components consist of at least three nodes. The simulations with the models use the parameters of the Hospital (Table \ref{tableTraces} and Appendix~\ref{sec:parameter_tuning}). In~(f-h) the results with the models are averages over 10 simulation runs. Results for all activity cycles, the Conference dataset, and for the simulated counterparts of the rest of the real networks are found in Appendices~\ref{sec:recurrent_components},~\ref{SecProps}. 
\label{FigMain}}
\end{figure*}

However, it has been recently revealed that face-to-face interaction networks exhibit structural and dynamical properties such as community formation, which originate from motion patterns that are far from random~\cite{SuneRecurrent}. In a temporal setting, communities are dynamic, meaning that their structure and size change over time. A common strategy to track dynamic communities is to construct their evolution timelines by aggregating connected components of at least three nodes in different time slots, according to some similarity measure~\cite{SuneRecurrent, GreeneComDet}. In other words, the building blocks of dynamic communities are connected components that appear recurrently. If we extract the connected components in each time slot of a real face-to-face interaction network, we can see that many of the exact same components appear several times throughout the observation period. Indeed, in Figs.~\ref{FigMain}a-c we have extracted and assigned IDs, in order of appearance, to the unique components found in three real-world datasets from SocioPatterns~\cite{SocioPatterns}: a Hospital, a Primary School and a High School \cite{HospitalData, PrimaryData, HighSchoolData} (see Table~\ref{tableTraces} and Appendices~\ref{sec:datasets},~\ref{sec:recurrent_components}, where we also consider a fourth dataset from a conference~\cite{ConferenceData}). The blue lines in Figs.~\ref{FigMain}a-c represent \emph{recurrent components}, i.e., components that appeared at least once in a previous time interval. By contrast, in the attractiveness model we observe very few recurrent components (Fig.~\ref{FigMain}d and Appendix~\ref{sec:recurrent_components}), even though the model accurately reproduces the broad distributions of contact durations and of times between consecutive contacts (Figs.~\ref{FigMain}f,g). This is because in the model nodes drift according to their own random trajectories and the probability for a group of at least three nodes to meet again is vanishing. In other words, components form in this model purely based on chance.

\begin{table}
\begin{tabular}{|c|c|c|c|c|c|c|c|c|}
\hline 
Dataset & $N$ & $T$ & $\bar{n}$ & $\bar{l}$ & Cycles & $\mu_1$ & $F_0$ & $\mu_2$\\ 
\hline 
Hospital & 70 & 4400 & 7.09 & 4.7 & 4 & 0.8 & 0.12 & 0.9\\ 
\hline 
Primary School & 242 & 3100 & 56.38 & 40.57 & 2& 0.35 & 0.2 & 0.78\\ 
\hline 
High School & 327 & 7375 & 41.89 & 25.56 & 5& 1.2 & 0.11 &0.86\\ 
\hline 
Conference & 113 & 7030 & 4.98 & 2.96 & 3 & 2.65 & 0.02 & 3.6\\ 
\hline 
\end{tabular}
\caption{Analyzed datasets. $N$ is the total number of agents; $T$ is the total duration of the dataset in slots of 20 seconds; $\bar{n}$, $\bar{l}$ are the average numbers of interacting agents and links (interactions) per slot. The activity cycles correspond to observation periods in different days (see Appendix~\ref{sec:datasets}). $\mu_1, F_0, \mu_2$ are the FDM parameters used in the simulated counterpart of each real network (see text).
\label{tableTraces}}
\end{table}

Here we present a model of mobile agents where their motion is not totally random, but instead it is also directed by pairwise similarity forces. We show that this model can capture the most distinctive features of face-to-face interaction networks including their observed recurrent component patterns. In addition to the two-dimensional Euclidean space where agents move and interact (an $L \times L$ square), agents in the model also reside in a hidden similarity space, where coordinates abstract their similarity attributes. Distances between the agents in this space act as similarity forces directing their motion towards other agents in the physical space and determining the duration of their interactions. We consider the simplest metric space as the similarity space, which is a circle of radius $R=N/2\pi$ where each agent $i=1, 2, \ldots, N$ is assigned a random angular coordinate $\theta_i \in [0, 2\pi]$. Therefore the similarity distance between two agents $i, j$ is $s_{ij} = R\Delta\theta_{ij}$, where $\Delta\theta_{ij} = \pi - \vert\pi-\vert\theta_i - \theta_j\vert\vert$ is the angular distance between the agents. (We also consider non-uniformly distributed coordinates in Appendix~\ref{sec:non_uniform}, obtaining similar results.)

Time in the model is slotted and at the beginning of each slot agents can be in one of two states: \emph{inactive} or \emph{interacting}. Inactive agents move in the slot only if they become active, while interacting agents move only if they escape their interactions. At the beginning of each slot $t$, each inactive agent $i$ is activated with a preassigned probability $r_i$. Furthermore, each interacting agent $i$ escapes its interactions with probability
\begin{equation}
\label{eq:escape_prob}
P_i^e(t) = 1 - \frac{1}{\vert \mathcal{N}_i(t)\vert} \sum_{j\in \mathcal{N}_i(t)} e^{-s_{ij}/\mu_1},
\end{equation}
where $\mathcal{N}_i(t)$ is the set of agents that $i$ is currently interacting with and $s_{ij}$ is the similarity distance between agents $i$ and $j$. The summands in Eq.~(\ref{eq:escape_prob}) can be seen as \emph{bonding forces} that decrease exponentially with the similarity distance, while parameter $\mu_1 > 0$ is the decay constant controlling the importance of these forces as the similarity distance increases and allowing us to tune the average contact duration (Appendix~\ref{sec:parameter_tuning}). The model assumes that the contact duration in number of slots between two agents $i, j$ is exponentially distributed with rate $s_{ij}/\mu_1$. The discrete analogue of this distribution is the geometric distribution with success probability $p_{ij}=1-e^{-s_{ij}/\mu_1}$. Therefore, Eq.~(\ref{eq:escape_prob}) is the average of $p_{ij}, j\in \mathcal{N}_i(t)$.

Each moving agent $i$ in the slot updates its position ($x^t_i$, $y^t_i$) according to the following motion equations
\begin{align}
\label{eq:motion_eqs_1}
x^{t+1}_i &= x^t_i + \sum_{j \in \mathcal{S}(t)}F_{ij}\frac{(x^t_j - x^t_i)}{\sqrt{(x^t_j - x^t_i)^2+(y^t_j - y^t_i)^2}} + R^x_i,\\
\label{eq:motion_eqs_2}
y^{t+1}_i &= y^t_i+ \sum_{j\in \mathcal{S}(t)} F_{ij}\frac{(y^t_j - y^t_i)}{\sqrt{(x^t_j - x^t_i)^2+(y^t_j - y^t_i)^2}} + R^y_i,
\end{align}
where $\mathcal{S}(t)$ is the set of all moving and interacting agents in the slot, while $F_{ij}$ is the magnitude of the \emph{attractive force} between agents $i$ and $j$, which also decreases exponentially with their similarity distance,
\begin{equation}
\label{eq:ForceEq}
F_{ij} = F_0 e^{-s_{ij}/\mu_2}.
\end{equation}
Parameter $F_0 \geq 0$ is the force magnitude at the minimum similarity distance, $s_{ij}=0$, while $\mu_2 > 0$ is the decay constant controlling the importance of the force magnitude as the similarity distance increases. Therefore, the sums in Eqs.~(\ref{eq:motion_eqs_1}), (\ref{eq:motion_eqs_2}) are the total attractive forces exerted to agent $i$ by the agents $j \in \mathcal{S}(t)$ along the $x$ and $y$ directions of the motion. The random motion components are $R^x_i= v\cos\phi_i$, $R^y_i = v\sin\phi_i$, where $\phi_i$ is  sampled uniformly at random from $[0, 2\pi$] and $v \geq 0$ is the magnitude of the random displacement. We can think of $R^x_i,  R^y_i$ as accounting for omitted degrees of freedom, akin to Langevin dynamics~\cite{schlick2010molecular}. At $v=0$ the motion becomes deterministic, while at $F_0=0$ it degenerates to random walks. Once the moving agents update their positions they either transition to the interacting state if they are within interaction range $d$ from other non-inactive agents, or to the inactive state. We call the described model \emph{Force-Directed Motion (FDM)} model. We make its implementation available at \cite{SimulatorURL}.

\begin{figure}[t]
\includegraphics[width=8.7cm]{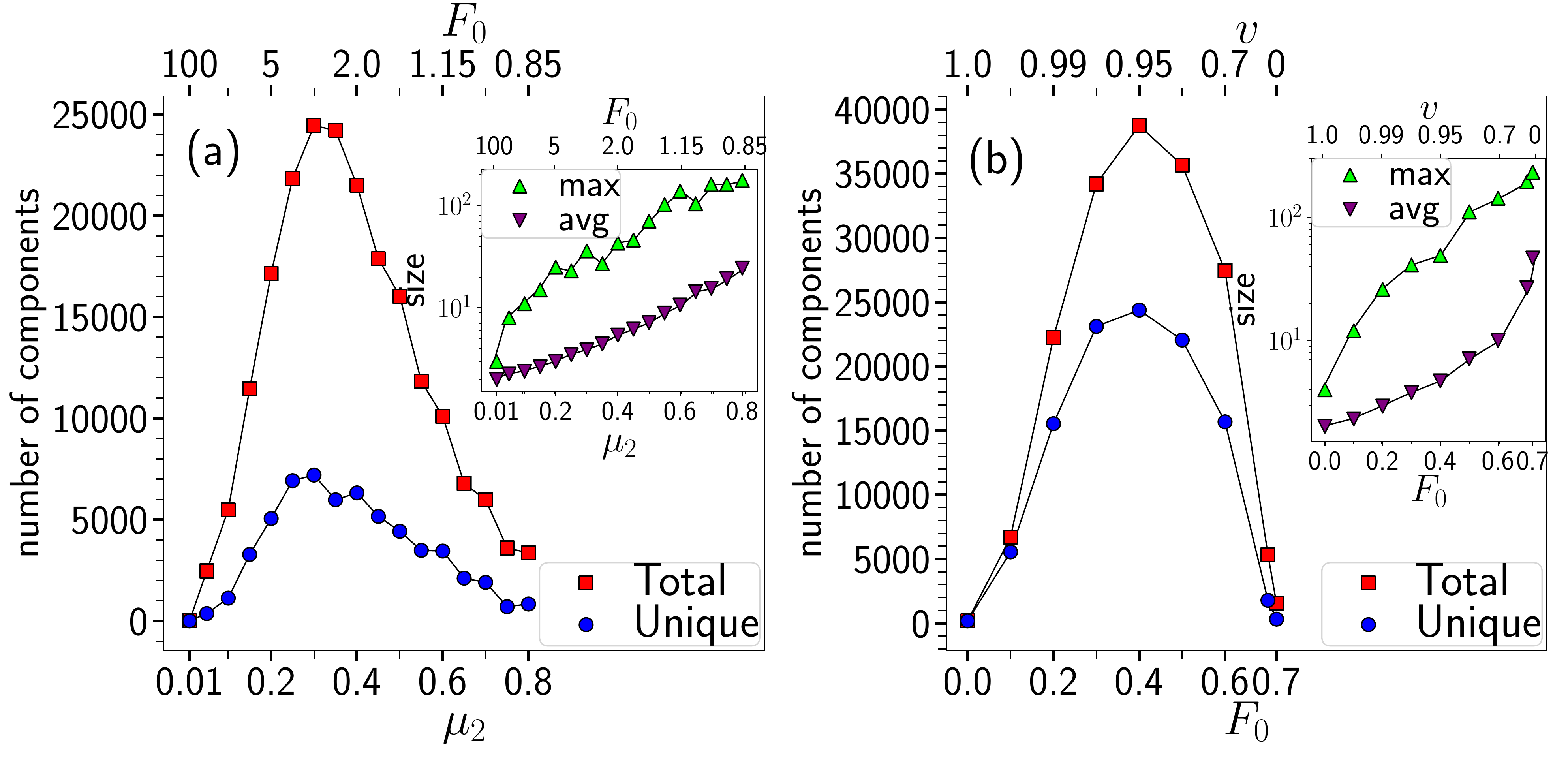}
\caption{Formation of components in the FDM. \textbf{(a)}~Number of components formed (total and unique) in deterministic motion ($v=0$) for pairs of parameters $ \mu_2$ (bottom $x$-axis) and $F_0$ (top $x$-axis). \textbf{(b)} Same as~(a) but for pairs of $F_0$ and $v \geq 0$. In both (a, b) as one parameter increases the other decreases so that the expected agent displacement per slot is always $\approx d=1$. The insets show the maximum and average size across all components. In both plots $N=242$, in (b)~$\mu_2=1$. See also Appendix~\ref{sec:parameter_tuning}.
\label{FigComponents}}
\end{figure}

To understand how the formation of components depends on $F_0, \mu_2, v$, we first consider deterministic motion. In this case, the magnitude of the expected agent displacement is controlled by $F_0$ and $\mu_2$. This magnitude can be kept fixed if, when $F_0$ decreases, $\mu_2$ increases accordingly. As $\mu_2$ increases, larger components form that involve agents at larger similarity distances, until the agents eventually collapse into a giant component. At the same time, the number of components initially increases and then decreases, see Fig.~\ref{FigComponents}(a). The motion in Eqs.~(\ref{eq:motion_eqs_1}),~(\ref{eq:motion_eqs_2}) is deterministic motion with random noise. This noise decreases the chances for similar---close in the similarity space---agents to meet, which reduces the size of components. At the same time, it can either increase (if its magnitude $v$ is sufficiently small) or decrease (if $v$ is sufficiently large) the number of components (Fig.~\ref{FigComponents}(b)).

\begin{figure*}[t]
\includegraphics[width=18cm]{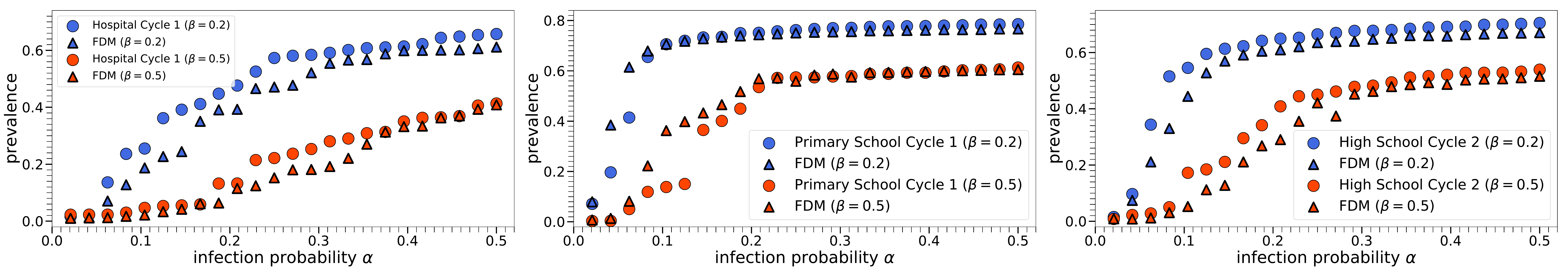}
\caption{Average percentage of infected agents per time slot (prevalence) of the SIS model as a function of the infection probability $\alpha$ in real and simulated networks (circles and triangles respectively), for two recovery probabilities $\beta$.  In the SIS each agent can be in one of two states, susceptible or infected. At any time slot an infected agent recovers with probability $\beta$ and becomes susceptible again, whereas infected agents infect the susceptible agents with whom they interact, with probability $\alpha$. To simulate the SIS process on temporal networks we use the dynamic SIS implementation of the Network Diffusion Library~\cite{NDlib}. See Appendix~\ref{sec:sis} for further details.
\label{FigSpread}}
\end{figure*}

To tune FDM's parameters in simulations of real networks we follow the procedure in Appendix~\ref{sec:parameter_tuning}. In a nutshell, we fix $v=d=1$. The number of agents $N$ and time slots $T$ are the same as in the real networks (Table~\ref{tableTraces}). The activation probability $r_i$ is either $r_i=0.5$ for every agent $i$ (Primary and High School), or sampled uniformly at random from $[0,1]$. Parameters $\mu_1, F_0, \mu_2$ (Table~\ref{tableTraces}) and the size of the Euclidean space $L$ (Appendix Table~\ref{tableParameters}) are adjusted in order to approximately match the following quantities between simulated and real networks: (i) the average contact duration (using $\mu_1$); (ii) the average number of recurrent components per interval of $10$ minutes, while ensuring a similar size of the largest component formed (using $F_0, \mu_2$); and (iii) the average agent degree in the time-aggregated network (using $L$).

\begin{figure}[t]
\includegraphics[width=8.72cm]{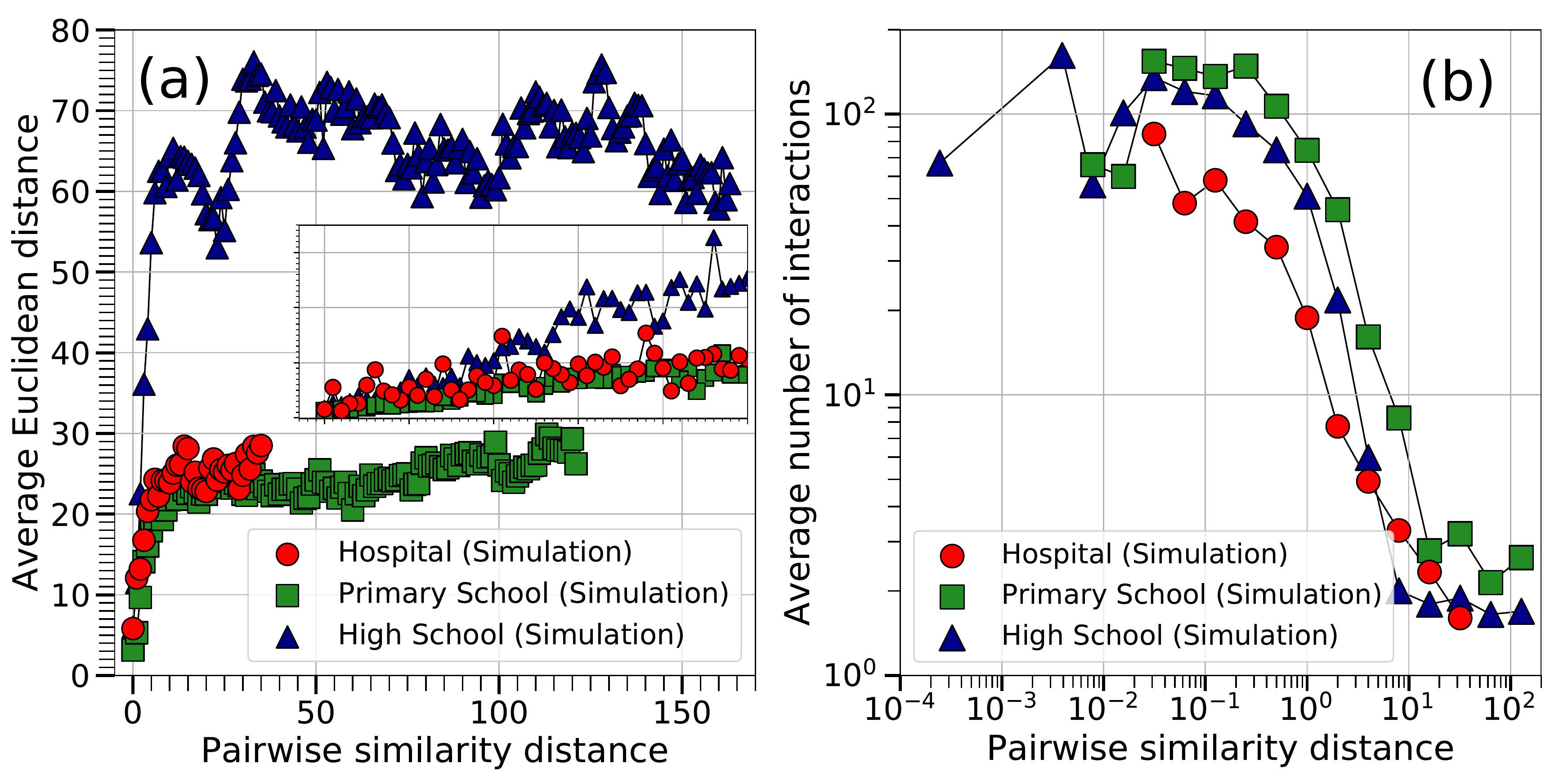}
\caption{Average Euclidean distance and number of interactions between two agents as a function of their similarity distance, in simulated counterparts of the Hospital, Primary School and High School. The inset in~(a) is a zoom in on similarity distances up to $5$.
\label{FigSPACES}}
\end{figure}

In Fig.~\ref{FigMain}e we see that the FDM can reproduce a similar pattern of unique and recurrent components as in the Hospital (Fig.~\ref{FigMain}a), in stark contrast to the attractiveness model (Fig.~\ref{FigMain}d). Similar results hold for all cycles of activity and for all considered datasets (Appendix~\ref{sec:recurrent_components}). In Fig.~\ref{FigMain}(h) we also see that the model can capture the correlations between the average number of recurrent components where an agent participated and the total number of interactions of the agent (see also Appendix~\ref{SecINVREC}). At the same time, the model reproduces the broad distributions of contact durations and of times between consecutive contacts (Figs.~\ref{FigMain}f,g). The model also adequately reproduces a range of other properties of the considered real networks, including weight distributions, distributions of component sizes and of shortest time-respecting paths, and group interaction durations (Appendix~\ref{SecProps}). It is then not surprising that the susceptible-infected-susceptible (SIS) spreading process~\cite{sis_ref} behaves similarly in real and  simulated networks (Fig.~\ref{FigSpread}). Fig.~\ref{FigSPACES} shows that agents close in the similarity space tend to stay closer to each other in the Euclidean space throughout the simulations and interact more often, as expected.

The exponential form of the attractive force in Eq.~(\ref{eq:ForceEq}) promotes locality and the formation of small components, as observed in real data. This is also promoted by the metric property of the similarity space, i.e., the triangle inequality, which ensures that if an agent $a$ is similar to an agent $b$ and $b$ is similar to a third agent $c$, then $c$ is also similar to $a$. This means that these agents will tend to gather close to each other in the Euclidean space forming triangle $abc$. On the other hand, if similarity distances do not satisfy the triangle inequality, then agents $a$ and $c$ might be close to some other agents $d$~and~$e$, forming chain $dabce$ in the network. In other words, agents will tend to form larger components. We verify this argument in Appendix~\ref{sec:metric_space}, where we break the triangle inequality by randomly assigning similarity distances to all pairs of agents instead of assigning to the agents similarity coordinates. In this way forces lose their localization effect and we see that a giant connected component, non-existent when the similarity space satisfies the metric property, forms in the middle of the Euclidean space.

In summary, forces emerging from similarity distances in metric spaces appear to provide a natural explanation for the observed recurrent component dynamics in face-to-face interaction networks. These forces direct the motion of the agents in the physical space and determine the agents' interaction durations. Motion based on these principles can still capture a wide range of other main properties of such networks, in addition to their recurrent component patterns. The interactions do not have to be exactly face-to-face or of few activity cycles. In Appendices~\ref{sec:recurrent_components},~\ref{SecProps} we see that similar results hold in a longitudinal  dataset from an MIT dormitory, where proximity was captured if mobile phones were within 10 meters from each other~\cite{DongMIT}.

The modeling approach we consider bears similarities to $N$-body simulations and Langevin dynamics~\cite{schlick2010molecular}, suggesting that similar techniques and approaches from these well established areas of physics can be applicable to contemporary network science problems. Yet, we note that the similarity forces in our case only direct the motion of the agents in the physical space, and do not depend on the agents' distances in this space akin to gravity. 

We also observe that \emph{hyperbolic} spaces appear to underlie the topologies of traditional complex networks, whose degree distributions are heterogeneous~\cite{Krioukov2010}. In this case, the hidden distance between two nodes is not  just the angular distance $R\Delta\theta$ but the effective distance $\chi=R\Delta\theta/(\kappa \kappa')$, where $\kappa, \kappa'$ are the expected degrees of the nodes~\cite{Krioukov2010}. One can replace angular with effective distances in the FDM. However, in all datasets we considered, the distribution of $\kappa$s was quite homogeneous to justify the need for this description~\footnote{an agent's $\kappa$ is its average degree per time slot}. Indeed, if we use effective distances in the FDM with the estimated $\kappa$s from the real data we obtain very similar results (Appendix~\ref{sec:hyperbolic}).

A natural direction for future work is the \emph{inverse problem} of inferring the similarity coordinates of agents given a sequence of real network snapshots. Another direction is extending the model with the addition of static nodes that exist both in the physical and in the similarity space and represent locations. Finally, it would be interesting to investigate how social influence could also be incorporated into the model, where interacting agents may influence each other and become more similar~\cite{leenders1997}. This would result in the agents moving both in the Euclidean and similarity spaces. Taken altogether, our results pave the way towards more realistic modeling of face-to-face interaction networks, which is crucial for understanding and predicting social group dynamics and designing efficient epidemic control and navigation~\cite{temporal_navigation, Boguna2008} strategies.
 
\begin{acknowledgments}
The authors acknowledge support by the EU H2020 NOTRE project (grant 692058).
\end{acknowledgments}

\onecolumngrid

\appendix

\section{Datasets}
\label{sec:datasets}

The considered real-world data are obtained from the SocioPatterns collaboration~\cite{SocioPatterns} and the MIT Social Evolution experiment~\cite{DongMIT}. The SocioPatterns data correspond to the face-to-face interaction networks of: (i) a Hospital ward in Lyon~\cite{HospitalData}; (ii) a Primary School in Lyon~\cite{PrimaryData}; (iii) a High School in Marseilles~\cite{HighSchoolData}; and (iv) a scientific Conference (Hypertext 2009) in Turin~\cite{ConferenceData}. The data were collected through the use of Radio-Frequency Identification (RFID) badges worn by individuals. Interactions were detected only if the badges were within $1$-$1.5$ meters in front of each other and exchanged at least $1$ radio packet in a $20$ seconds interval. Therefore each time slot in the data has duration $20$ seconds and corresponds to a network snapshot. The MIT Social Evolution dataset contains a record of the proximity of students in a student hall of MIT, captured with the Bluetooth capabilities of the students' mobile phones. The phones could detect proximity of other phones within a radius of $10$ meters in all directions, including different floors. The resolution of this dataset is $6$ minutes, which is the frequency by which the phones emitted a Bluetooth signal to be detected by other phones nearby~\cite{DongMIT}.

The interactions in the Hospital were collected during a period of $5$~days (December $6$-$10$, 2010) and involve $N=75$ nodes ($29$ patients and $46$ health-care workers); in the Primary School they were collected in $2$~days (October $1^{\text{st}}$, $2^{\text{nd}}$, 2009) and involve $N=242$ nodes ($232$ children and $10$ teachers); in the High School in $5$~days (December $2$-$6$, 2013) and involve $N=327$ nodes (students); in the Conference over $2.5$~days (June $29^{\text{th}}$ to July $1^{\text{st}}$, 2009) and involve $N=113$ nodes (participants); and in the MIT Social Evolution the data were collected during a period of $8$ months (October 2008 - May 2009) and involve $N = 74$ nodes (students). We have considered the periods described below. 

\begin{itemize}
\item[(i)] \underline{Hospital}. In the Hospital there are two working shifts, a morning-afternoon shift and an afternoon-night shift. Health-care workers that are present in one shift are usually not present in the other shift. We have considered the four morning-afternoon shifts from 7am to 1:30pm on December $7^\textnormal{th}$-$10^\textnormal{th}$, which correspond to activity cycles~$1$-$4$. These cycles have a total duration of $4400$ time slots involving $N=70$ nodes. Each cycle has a duration of $1100$ slots, beginning at the earliest recorded interaction in the corresponding observation day. There are $43$-$46$ nodes present in each activity cycle out of the $70$ nodes observed in the total duration of the dataset.
 
\item[(ii)] \underline{Primary School}. For each of the two observation days we have considered the working periods from 8:30am to 4:30pm \cite{PrimaryData}. These two periods correspond to activity cycles~1 and~2 and have a total duration of $3100$ slots. Cycle~1 has duration of $1555$ slots and consists of  $238$ nodes, while cycle~2 has duration of $1545$ slots and consists of $236$ nodes out of the $242$ nodes observed in the total duration of the dataset.

\item[(iii)] \underline{High School}. Here the working hours are not explicitly stated. We use the snapshot timestamps in the data to identify the periods of activity in each of the five observation days, as the first timestamp in each day is several hours ahead of the last timestamp in the previous day~\cite{SocioPatterns}. Activity cycle~1 has duration $899$ slots, while each of the activity cycles 2-5 has duration $1619$ slots. There are $295$-$312$ nodes present in each activity cycle out of the $327$ nodes observed in the total duration of the dataset ($7375$ slots.)

\item [(iv)] \underline{Conference}. We have identified the cycles of activity in each of the three observation days using the snapshot timestamps as in the High School dataset. Activity cycles~1, 2, 3 have durations $2874, 2210, 1946$ slots, respectively. There are $97$-$102$ nodes present in each activity cycle out of the $113$ nodes observed in the total duration of the dataset ($7030$ slots).

\item [(v)] \underline{MIT Social Evolution}. Finally, in this dataset we have considered only the proximities detected with high probability in the same floor in the period from October 2008 to May 2009. Specifically, in the data each detected proximity has a timestamp and a field called ``prob2'', which is the probability that the two phones detecting proximity were on the same floor. We considered only proximities with $\textnormal{prob2} \geq 0.5$ in time slots of $6$ minutes. This gives a total of $60905$ slots involving $62$ nodes. In this dataset there are no clearly identifiable activity cycles.

\end{itemize} 

We make the processed datasets available at \cite{SimulatorURL}.

\section{Recurrent components}
\label{sec:recurrent_components}

\subsection{Extraction process}

Given a real or simulated network we first find all connected components in each time slot of the network using the Disjoint Set Union algorithm from~\cite{GallerUnionFind}. Each identified component is a set of at least three nodes. We then go over all time slots from the beginning to the end and assign IDs $1, 2, \ldots,$ etc., to their components as follows. If a component is seen for the first time, i.e., it does not consist of exactly the same nodes as a component seen in a previous slot, it is assigned a new ID and it is marked as \emph{unique}; if more than one unique components are found in a slot they are assigned new IDs arbitrarily. Otherwise, if a component consists of exactly the same nodes as a previously seen component, it is assigned the ID of that component and it is marked as \emph{recurrent}. 

In Figs.~\ref{FigMain}a-e of the main text and Figs.~\ref{figHPComps}-\ref{figHYTComps} below, the observation period ($x$-axis) is binned into $30$ minute intervals, while in Fig.~\ref{figMITComps} the observation period is binned into $60$ minute intervals. The black lines spanning each interval indicate in the $y$-axis the IDs of the unique components found in the slots of the interval. Similarly, the blue lines indicate the IDs of the recurrent components found in each interval.

\subsection{Unique and recurrent components in real and modeled networks}

\begin{figure}[!htb]
\centering
\includegraphics[width=16cm,height=8cm]{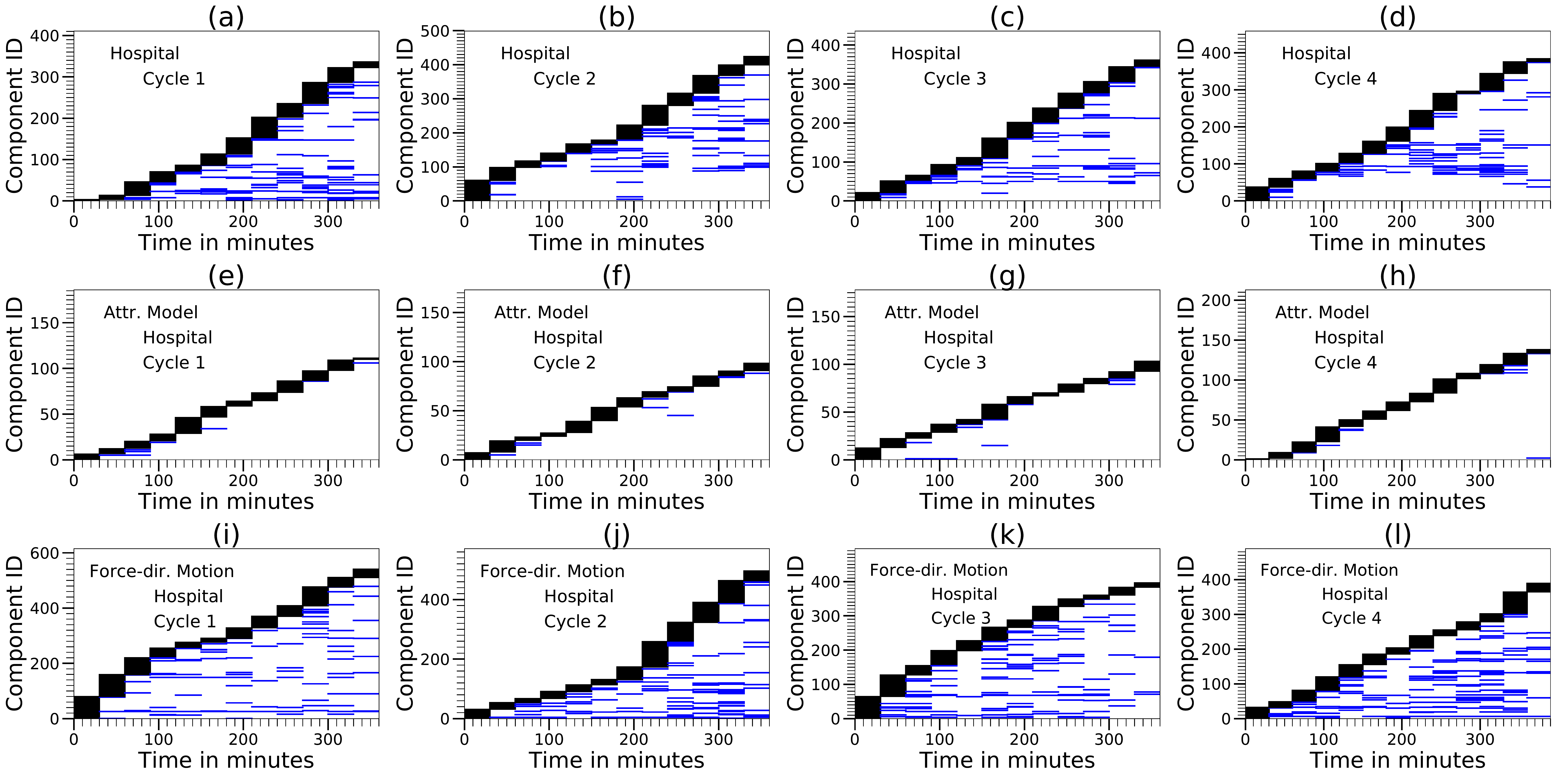}
\caption{\textbf{(a-d)}~Unique and recurrent components found in each cycle of activity in the Hospital. \textbf{(e-h)}~Components found in a simulation run of the attractiveness model assuming activity cycles of the same durations as in~(a-d). \textbf{(i-l)}~Same as~(e-h) but for the FDM (Force-dir.~Motion) model. All simulations use the Hospital parameters (Table~\ref{tableParameters} in Sec.~\ref{sec:parameter_tuning}). 
\label{figHPComps}}
\end{figure}

\begin{figure}[!htb]
\centering
\includegraphics[width=17.7cm, height=8cm]{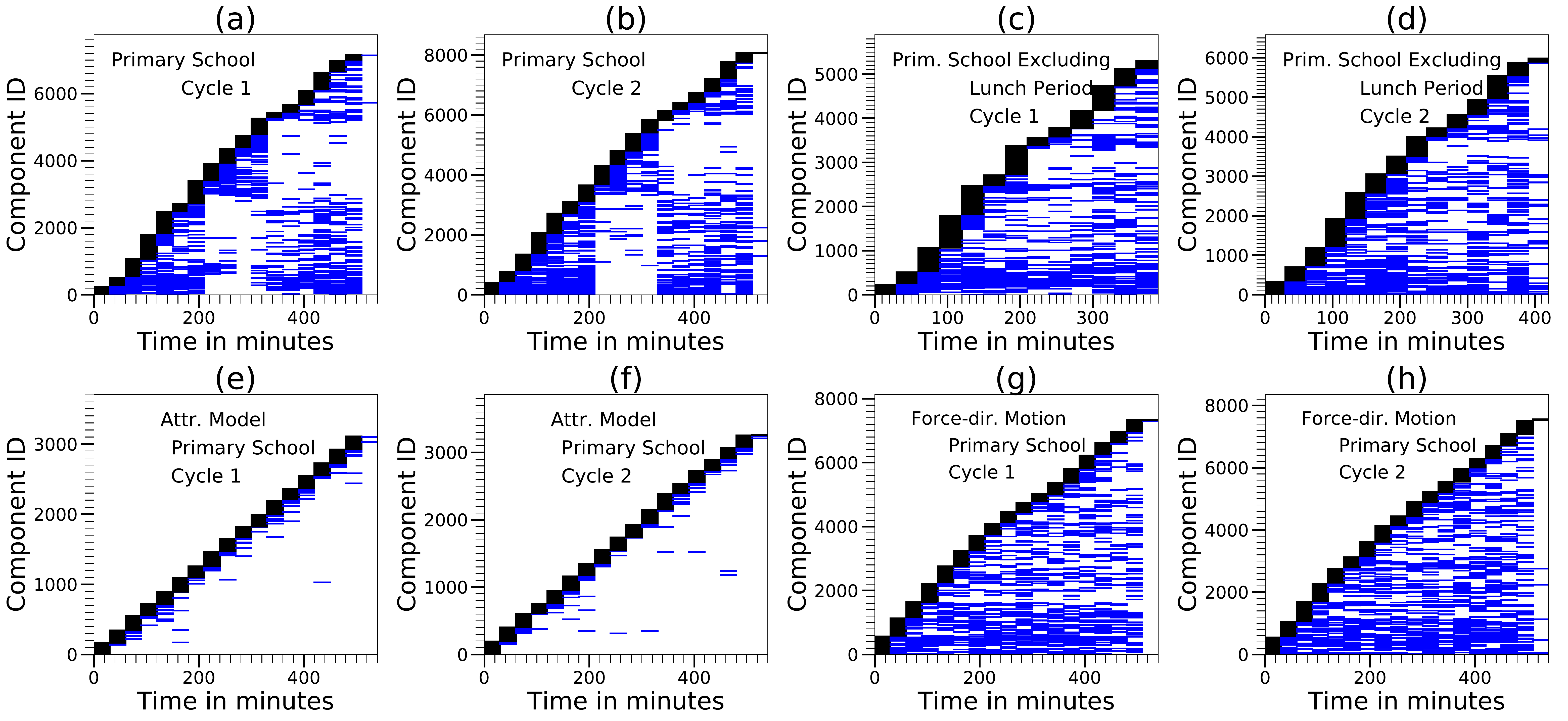}
\caption{\textbf{(a, b)}~Unique and recurrent components found in each cycle of activity in the Primary School. \textbf{(c, d)} Same as (a,~b) but excluding the lunch break period. \textbf{(e, f)}~Components found in a simulation run of the attractiveness model assuming activity cycles of the same durations as in~(a, b). \textbf{(g, h)}~Same as~(e, f) but for the FDM (Force-dir.~Motion) model. All simulations use the Primary School parameters (Table~\ref{tableParameters} in Sec.~\ref{sec:parameter_tuning}).
\label{figPSComps}}
\end{figure}

\begin{figure}[!htb]
\centering
\includegraphics[width=17.7cm, height=10cm]{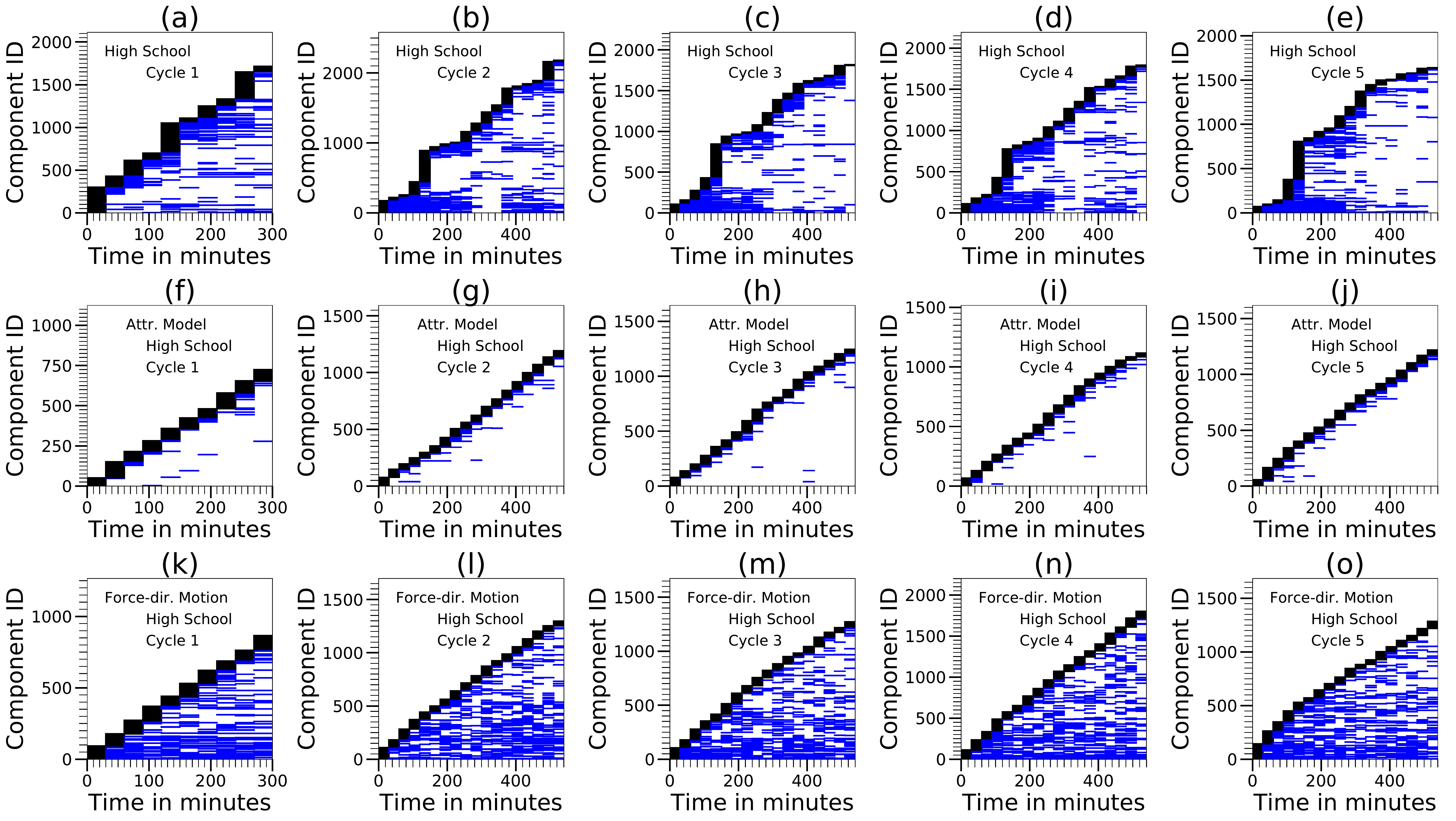}
\caption{\textbf{(a-e)}~Unique and recurrent components found in each cycle of activity in the High School. \textbf{(f-j)}~Components found in a simulation run of the attractiveness model assuming activity cycles of the same durations as in~(a-e). \textbf{(k-o)}~Same as~(f-j) but for the FDM (Force-dir.~Motion) model. All simulations use the High School parameters (Table~\ref{tableParameters} in Sec.~\ref{sec:parameter_tuning}). 
\label{figHSComps}}
\end{figure}

\begin{figure}[!htb]
\centering
\includegraphics[width=16cm,height=3.5cm]{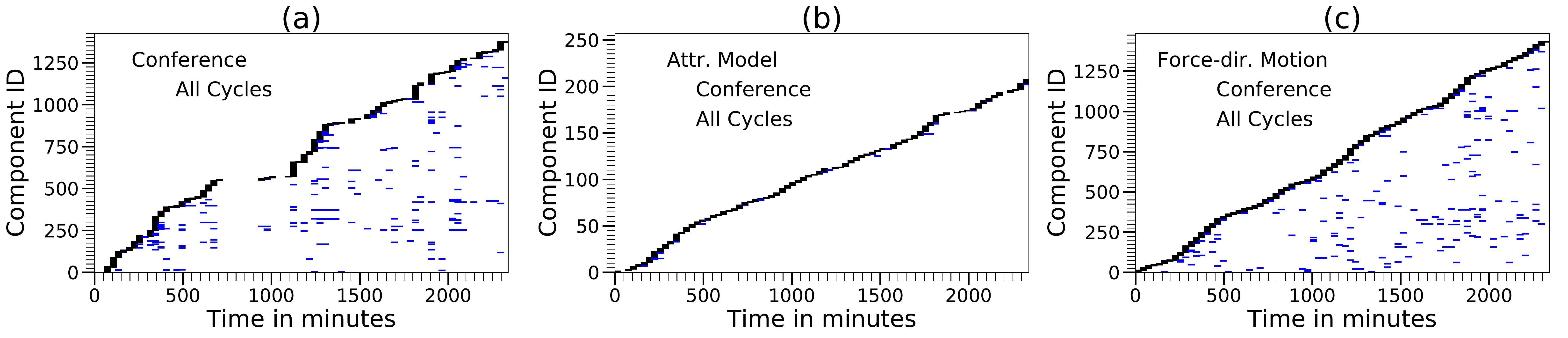}
\caption{\textbf{(a)} Unique and recurrent components found over the whole duration of the Conference dataset. \textbf{(b)}~Components found in a simulation run of the attractiveness model with the same duration as in~(a). \textbf{(c)}~Same as~(b) but for the FDM (Force-dir.~Motion) model. All simulations use the Conference parameters (Table~\ref{tableParameters} in Sec.~\ref{sec:parameter_tuning}).
\label{figHYTComps}}
\end{figure}

\begin{figure}[!htb]
\centering
\includegraphics[width=16cm,height=3.5cm]{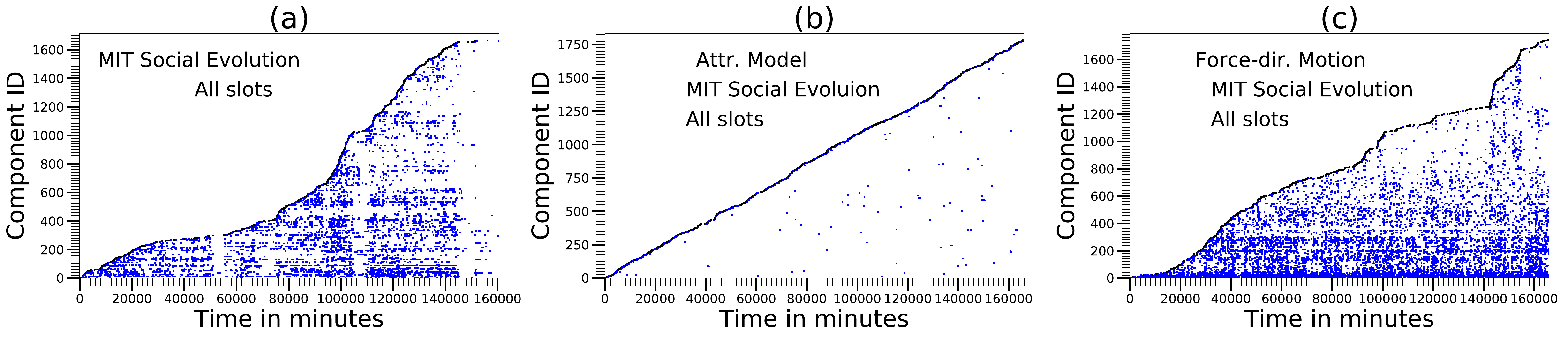}
\caption{\textbf{(a)} Unique and recurrent components found over the whole duration of the MIT Social Evolution dataset. \textbf{(b)}~Components found in a simulation run of the attractiveness model with the same duration as in~(a). \textbf{(c)}~Same as~(b) but for the FDM (Force-dir.~Motion) model. All simulations use the MIT Social Evolution parameters (Table~\ref{tableParameters} in Sec.~\ref{sec:parameter_tuning}).
\label{figMITComps}}
\end{figure}

Figs.~\ref{figHPComps}-\ref{figMITComps} show the unique and recurrent components in the real datasets and in corresponding simulated networks with the attractiveness and FDM models. For the Hospital, Primary School and High School the results are shown for each activity cycle. For the Conference the results are shown for the whole duration (all activity cycles), as there were relatively few recurrent components in each individual activity cycle. Furthermore, for the Primary School we also show the results if we exclude the lunch break period in each activity cycle (12pm-2pm) where children of different classes have lunch in a common place and some children go home to have lunch~\cite{PrimaryData}. Removing this period results in a more uniform pattern of recurrent components formation (Figs.~\ref{figPSComps}a,b vs. Figs.~\ref{figPSComps}c,d).

As in the main text, we see that in simulated networks with the attractiveness model recurrent components are almost non-existent, while they are abundant in simulated networks with the FDM as in the real data. We note that if attraction forces are disabled in the FDM ($F_0=0$), agents perform random walks, and the results are similar to the attractiveness model.

\subsection{Recurrent components and node interactions}
\label{SecINVREC}

\begin{figure}[!htb]
\centering
\includegraphics[width=15cm,height=8.5cm]{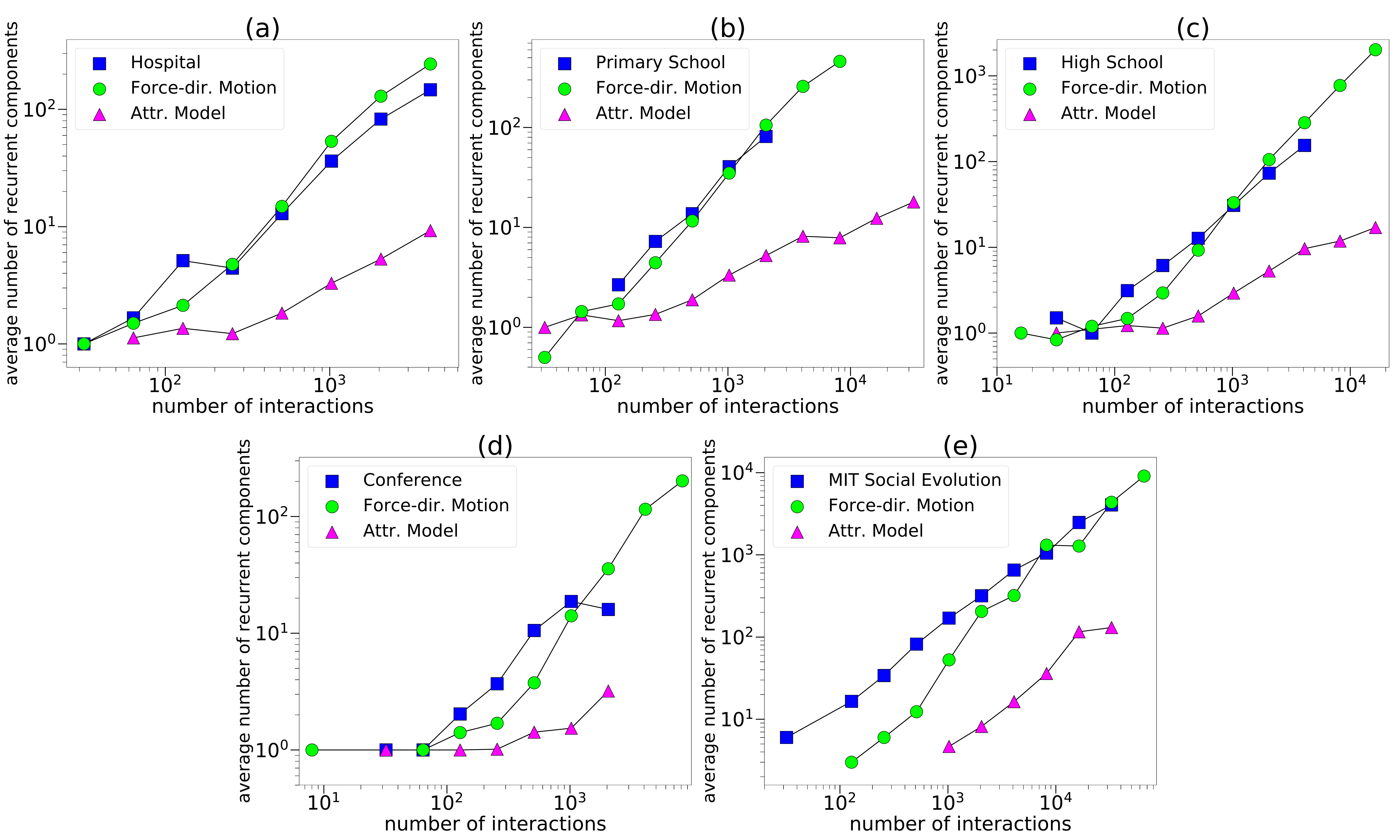}
\caption{Average number of recurrent components where a node participates as a function of the total number of interactions of the node in the datasets and in simulated networks with the FDM (Force-dir.~Motion) and attractiveness models. For each real dataset the corresponding simulations with the models use the  dataset's parameters (Sec.~\ref{sec:parameter_tuning}). The results in~(a-d) are averages over $10$ simulation runs, while (e) shows results from one simulation run. 
\label{IntVSRec}}
\end{figure}

Fig.~\ref{IntVSRec} (and~\ref{FigMain}h in the main text) shows the correlations between the average number of recurrent components where a node participates and its total number of interactions in the real datasets and in the corresponding simulated networks. The total number of interactions of a node $i$, $I_i$, is the total number of edges (interactions) between $i$ and other nodes $j \neq i$ over the duration of the dataset. This metric is the same as the \emph{strength} of the node in the time-aggregated network of contacts (Sec. \ref{SecProps}). The number of recurrent components where a node participates is the total number of such components where the node is a member of over the duration of the dataset. We measure the recurrent components within intervals of 30 minutes (blue lines in  Figs.~\ref{figHPComps}-\ref{figHYTComps}) and of 60 minutes (Fig.~\ref{figMITComps}). A recurrent component appearing more than once in an interval is counted only once. We see that the FDM can better capture the behavior in the real networks compared to the attractiveness model, as expected. We again note that if attraction forces are disabled in the FDM ($F_0=0$) the results are similar to the attractiveness model.

\section{Other properties of real versus modeled networks}
\label{SecProps}

\begin{figure}[!h]
\centering
\includegraphics[width=17.8cm,height=6.2cm]{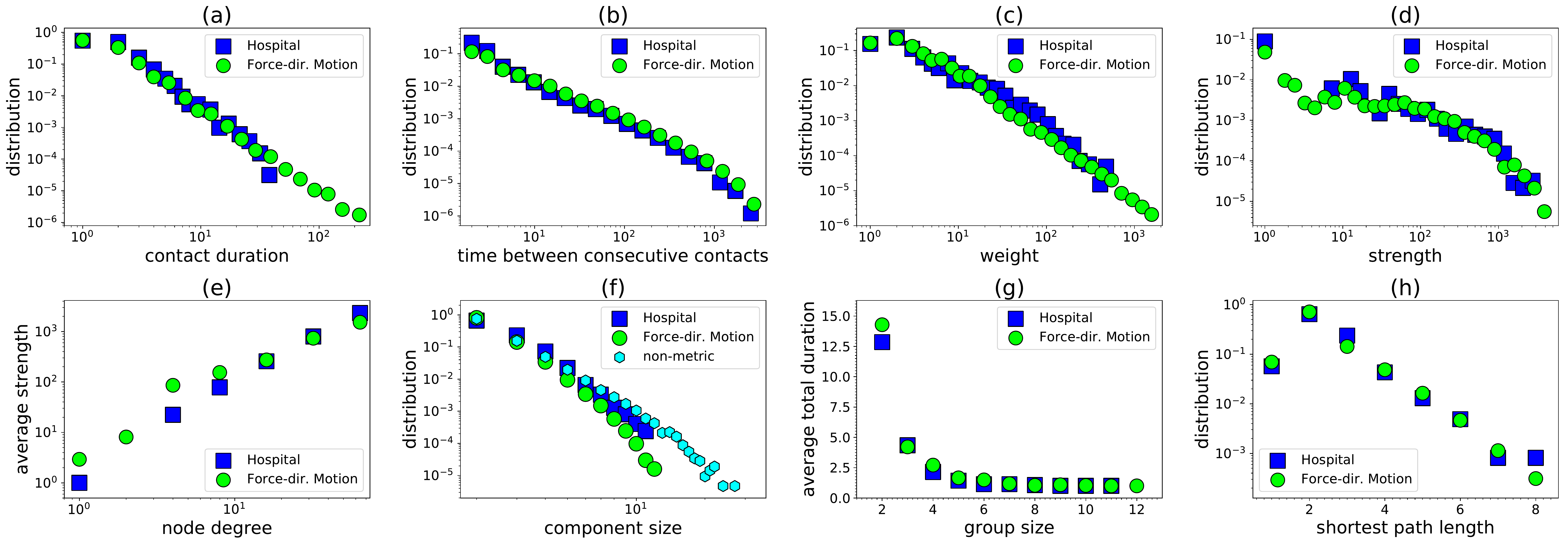}
\caption{Properties of the Hospital face-to-face interaction network and of corresponding simulated networks with the FDM (Force-dir.~Motion) model. 
\textbf{(a)}~Distribution of contact duration. 
\textbf{(b)}~Distribution of time between consecutive contacts. 
\textbf{(c)}~Weight distribution. 
\textbf{(d)}~Strength distribution. 
\textbf{(e)}~Average node strength as a function of node degree. 
\textbf{(f)}~Distribution of component sizes. 
\textbf{(g)}~Average total interaction duration of a group as a function of its size. 
\textbf{(h)}~Distribution of shortest time-respecting paths.
In all cases the simulation results are averages over $10$ runs. The distributions in (a)-(d) have been binned logarithmically; (e) also uses logarithmic binning. Plot (f) also shows the results if we randomly assign similarity distances to pairs of nodes (non-metric) instead of assigning to nodes similarity coordinates (see Sec.~\ref{sec:metric_space}).
\label{figHPProps}}
\end{figure}

\begin{figure}[!h]
\centering
\includegraphics[width=17.8cm,height=6.2cm]{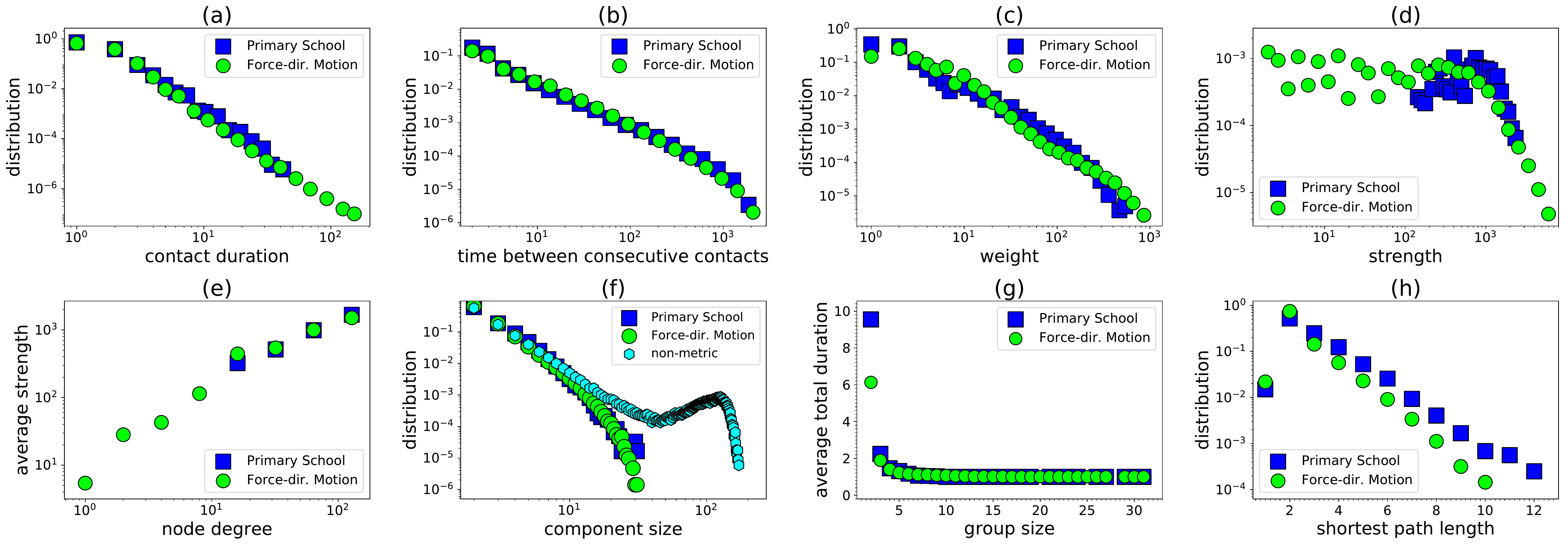}
\caption{Same as Fig.~\ref{figHPProps} but for the Primary School.
\label{figPSProps}}
\end{figure}

\begin{figure}[!h]
\centering
\includegraphics[width=17.8cm,height=6.2cm]{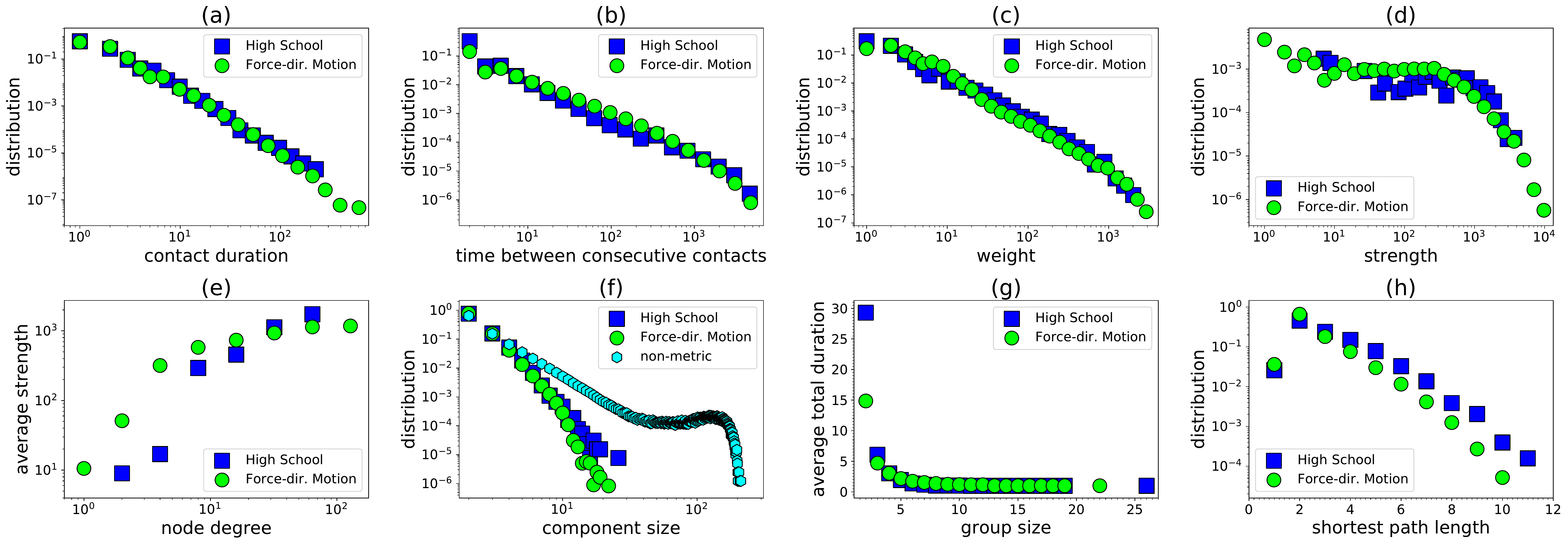}
\caption{Same as Fig.~\ref{figHPProps} but for the High School.
\label{figHSProps}}
\end{figure}

\begin{figure}[!h]
\centering
\includegraphics[width=17.8cm,height=6.2cm]{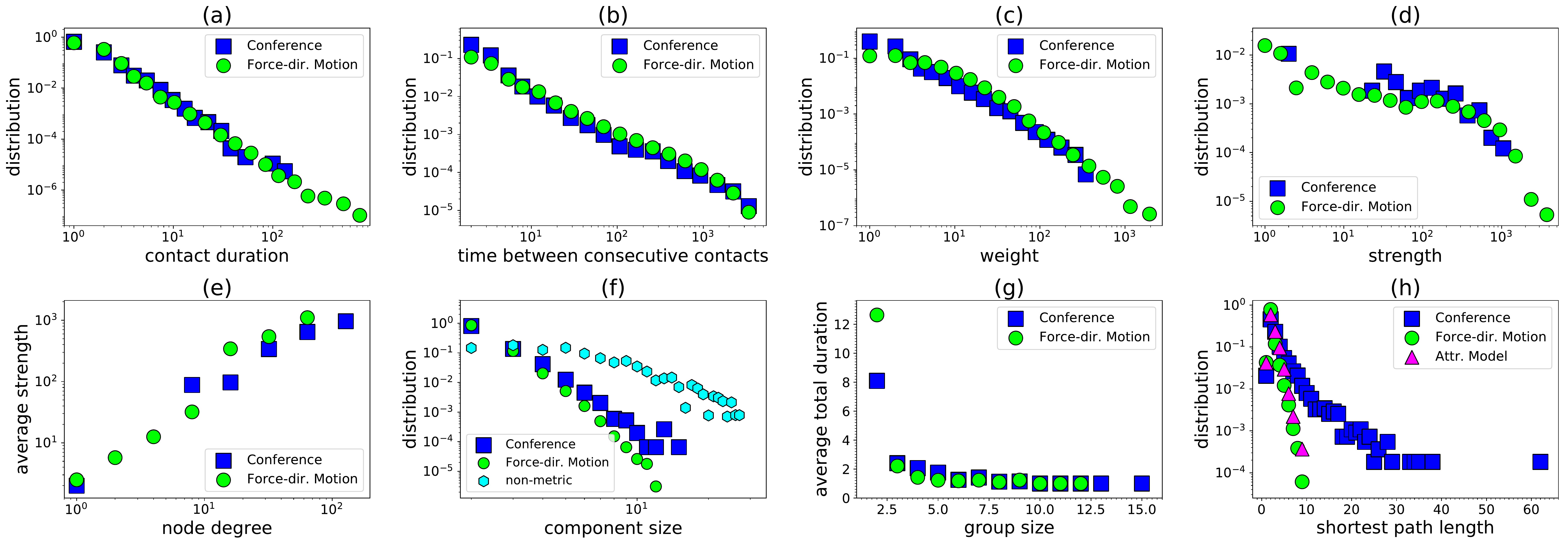}
\caption{Same as Fig.~\ref{figHPProps} but for the Conference. Plot (h) also shows the corresponding simulation results with the attractiveness model.
\label{figHYTProps}}
\end{figure}

\begin{figure}[!h]
\centering
\includegraphics[width=17.8cm,height=6.2cm]{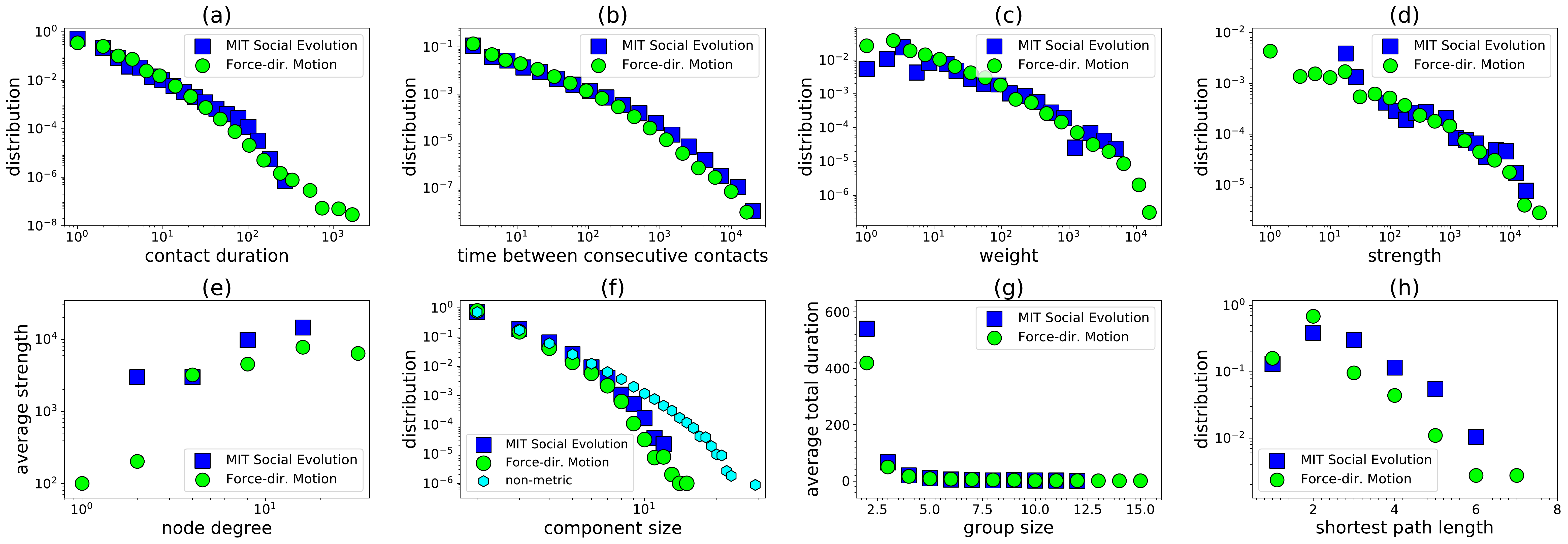}
\caption{Same as Fig.~\ref{figHPProps} but for the MIT Social Evolution. In all plots the simulation results are averages over 10 runs except from (h), which shows the results from one run as computing this metric in this large dataset is computationally expensive.
\label{figMITProps}}
\end{figure}

In Figs.~\ref{figHPProps}-\ref{figMITProps} we compare a range of other properties between the considered real networks and the corresponding simulated networks with the FDM model. Specifically, we consider the following eight properties computed over all time slots (these properties were also considered in~\cite{Starnini2016, Starnini2013}):

\begin{itemize}
\item[(a)] \emph{Distribution of contact duration}. This is the distribution of the time duration (in number of time slots) that two nodes remain in contact (interact). 

\item[(b)] \emph{Distribution of time between consecutive contacts}. This is the distribution of time (in number of time slots) that elapses between the last time that a pair of nodes interacted till the time that the same pair of nodes interacts again. 

\item[(c)] \emph{Weight distribution}. To compute this distribution we first construct the time-aggregated network of contacts, where two nodes are connected by an edge if they interacted at least once. Each edge has a weight equal to the total number of time slots that the corresponding nodes interacted. The weight distribution is the distribution of the edge weights.

\item [(d)] \emph{Strength distribution}. This is the distribution of node strengths. The strength of a node is the sum of the weights of all edges attached to the node in the time-aggregated network of contacts. 

\item[(e)] \emph{Average node strength as a function of node degree}. From the time-aggregated network of contacts we also compute the degree of each node (sum of edges attached to the node) and for each degree we compute the average strength among nodes with that specific degree.

\item[(f)] \emph{Distribution of component sizes}. This is the distribution of the number of nodes in the connected components formed throughout the observation time, including components of size $2$.

\item[(g)] \emph{Average total interaction duration of a group as a function of its size}. The total interaction duration of a group of nodes is the total number of time slots throughout the observation time where the exact same group of nodes formed a connected component. For each group size we compute the average of this duration among groups with that specific size.

\item[(h)] \emph{Distribution of shortest time-respecting paths}. Consider three nodes $i$, $k$ and $j$, where $i$ and $k$ interact at slot $t$ and $k$ and $j$ interact at slot $t' > t$. In this example, the time-respecting path between $i$ and $j$ is $i \to k \to j$ and has length $2$. The shortest time-respecting path between $i$ and $j$ is the shortest such path throughout the observation time. We consider the distribution of lengths of the shortest time-respecting paths among all pairs of nodes that we compute using the PATHPY library~\cite{Pathpy}.
\end{itemize}
In Figs.~\ref{figHPProps}-\ref{figMITProps} we see that the FDM can adequately capture the characteristics of the real-world networks. An exception is the distribution of the shortest time-respecting paths between the Conference and the model, where we observe a significant deviation (Fig.~\ref{figHYTProps}h). However, this deviation is not due to the attraction forces in the FDM, since as we see in Fig.~\ref{figHYTProps}h the attractiveness model also yields similar results to the FDM. In fact, we observe that this distribution is also very different between the Conference and the other datasets---as can be seen, in the Conference there are significantly longer paths. This difference might be due to the fact that interactions are less structured in this dataset, in the sense that participants move at will between different areas such as conference rooms, coffee break areas, etc.~\cite{ConferenceData}, which also justifies the fewer recurrent components in this dataset compared to the rest (cf. Figs.~\ref{figHPComps}-\ref{IntVSRec}).

\section{Model parameters}
\label{sec:parameter_tuning}

The FDM has the following six parameters: (i)~$N$, which is the number of agents to simulate; (ii)~$T$, which is the number of time slots to simulate; (iii)~$L$, which determines the area of the two-dimensional Euclidean space where agents move and interact (an $L \times L$ square); (iv)~$\mu_1$ in Eq.~(\ref{eq:escape_prob}) of the main text, which controls the average contact duration; and (v, vi)~$F_0, \mu_2$ in Eq.~(\ref{eq:ForceEq}) of the main text, which control the expected agent displacement due to attraction forces and the abundance and size of components (see Fig.~\ref{FigComponents} in the main text and the related discussion and Appendix~\ref{sec:displacement}). The interaction radius $d$ and magnitude of random displacement $v$ are fixed to $v=d=1$. One can fix $d$ to any other value with  $v=d$, which will result in a rescaling of the size of the Euclidean space $L$. We also note that the radius of the similarity space in the FDM, $R = N/2\pi$, is a dummy parameter in the sense that if $R$ changes one can rescale $\mu_1, \mu_2$ such that the bonding and attraction forces (Eqs.~(\ref{eq:escape_prob}), (\ref{eq:ForceEq}) in the main text) remain the same. Below we discuss how we tune the above parameters in the simulated counterparts of each real network---see Table~\ref{tableParameters} for their values.

\begin{table}[!htb]
\begin{tabular}{|c|c|c|c|c|c|c|c|}
\hline 
Network & $N$ & $T$ & $T_{\textnormal{warmup}}$ & $L$ & $\mu_1$ & $F_0$ & $\mu_2$ \\ 
\hline 
Hospital & 70 & 4400 & 2500 & 95 & 0.8 & 0.12 & 0.9 \\ 
\hline 
Primary School & 242 & 3100 & 2000 & 98  & 0.35 & 0.2 & 0.78 \\ 
\hline 
High School & 327 & 7375 & 6500 & 295 & 1.2 & 0.11 & 0.86\\ 
\hline 
Conference & 113 & 7030 & 6000 & 340 & 2.65 & 0.02 & 3.6 \\ 
\hline
MIT Social Evolution & 62 & 60905 & 10000 & 2200 & 1.9 & 0.1 & 1.03 \\ 
\hline 
\end{tabular}
\caption{FDM parameter values used in the simulated counterpart of each real network.
\label{tableParameters}}
\end{table}

\begin{figure}[!h]
\centering
\includegraphics[width=17cm]{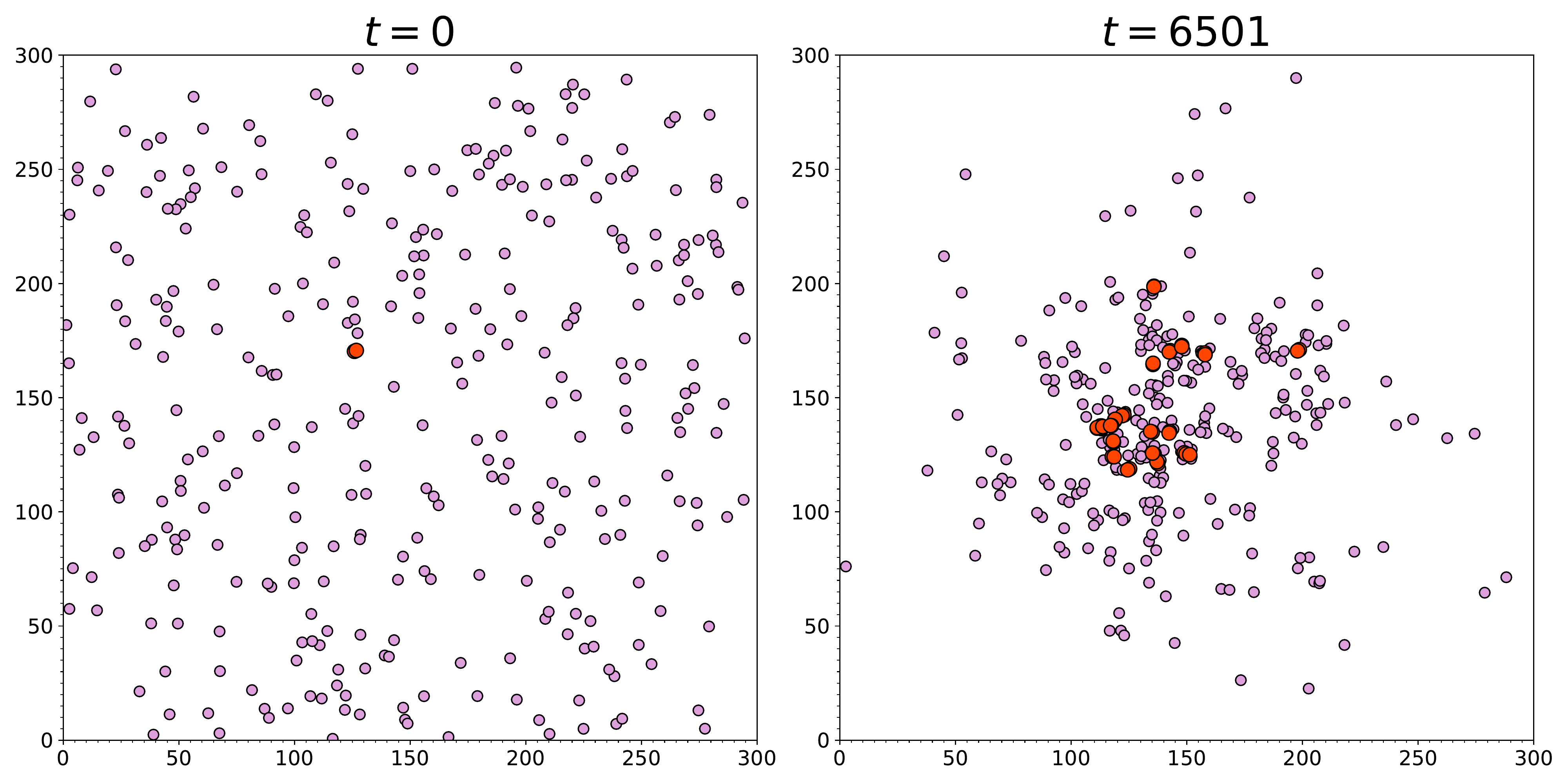}
\caption{Distribution of the agents in the Euclidean space at $t=0$ and $t=T_{\textnormal{warmup}}+1=6501$ in a simulated counterpart of the High School. Agents engaged in interactions are shown by red circles while the rest of the agents are shown by light purple circles.
\label{figInitialPos}}
\end{figure}

\begin{figure}[!h]
\centering
\includegraphics[width=15cm]{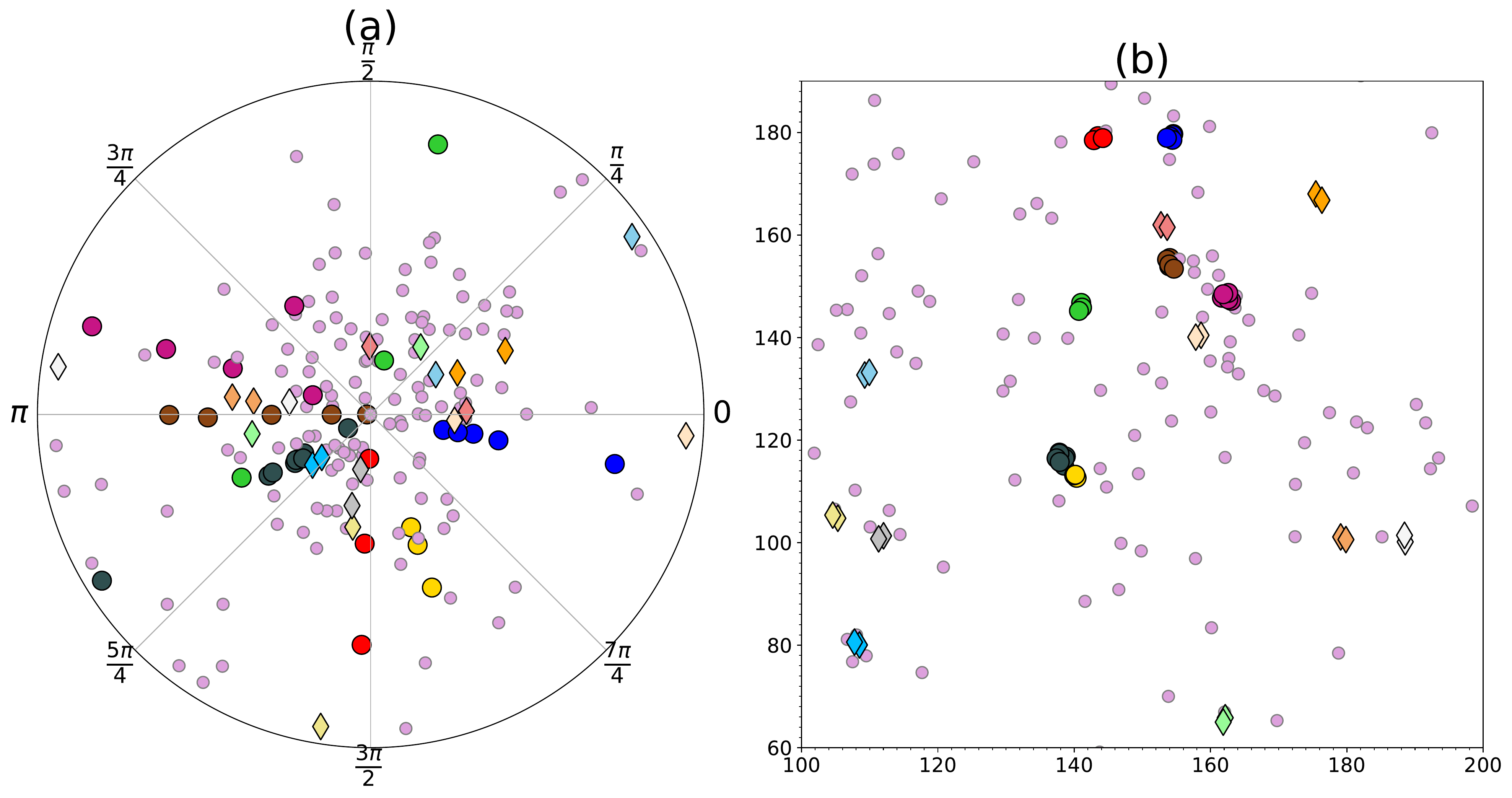}
\caption{Hidden similarity and physical Euclidean space of the agents. \textbf{(a)}~Similarity space of the agents in a simulated counterpart of the High School. The agents are colored, sized and marked as in the snapshot of their temporal network in~(b) and placed according to their angular (similarity) coordinates. For visualization purposes, the agents also have radial coordinates assigned using the formula $r_i = \mathcal{R} - \log{I_i}$, where $I_i$ is the total number of interactions of agent $i$ in the simulation, while $\mathcal{R} = \log{\max_i\{I_i\}}$ is the radius of the circle. \textbf{(b)}~Snapshot of the agents in the Euclidean space at time slot $t = 6108$ ($\textnormal{total slots}=7375$). The diamonds represent interactions involving only 2 agents, while the bigger circles represent interactions between at least 3 agents. The smallest grayed out circles are the moving agents that are not interacting, while inactive agents are not shown.
\label{figS1vsSnap}}
\end{figure}

Parameters $N, T$ are set equal to the total number of agents and time slots in the real dataset. $T_{\textnormal{warmup}}$ is a simulation warmup period until the average number of interacting agents per slot stabilizes.  All properties of the simulated networks are measured after this period. This period is required in order to give time to the agents that are close in the similarity space to move close to each other in the Euclidean space, as agents are initially uniformly distributed in the Euclidean space. One can avoid using a warmup period by assigning to the agents initial positions in the Euclidean space not uniformly at random but from a snapshot of a previous simulation run after $T_{\textnormal{warmup}}$, along with the similarity coordinates that the agents had in the run. We implement this option in our code that we make available at \cite{SimulatorURL}. Fig.~\ref{figInitialPos} shows snapshots of the agents in the Euclidean space at times $t=0$ and $t=T_{\textnormal{warmup}}+1$ in a simulated counterpart of the High School. Fig.~\ref{figS1vsSnap}b  also visualizes the agents in the Euclidean space at time $t = 6108$ after $T_{\textnormal{warmup}}$, while Fig.~\ref{figS1vsSnap}a shows the agents in their similarity space. As expected, we see that the majority of agents that participate in interactions in the snapshot are very close to each other in the similarity space along the angular direction.

For setting $L$, $\mu_1$, $F_0$, $\mu_2$ we follow a two-stage procedure that consists of a parameter initialization and a parameter tuning phase. We describe these two phases below.

\subsection{Parameter initialization}

\begin{itemize}
\item Parameter $L$: We set the initial value of this parameter based on the average number of interacting agents per slot $\bar{n}$ and the total number of agents $N$ in the dataset (see Table~\ref{tableTraces} in the main text for the values of $\bar{n}$). Specifically, assuming that there are no boundary effects, no inactive agents, a uniform spatial distribution of agents with density $\delta=N/L^2$, and an interaction radius $d=1$, the expected degree of an agent is $\bar{k} \approx N\pi /L^2$. Therefore, the probability that a given agent interacts with another agent is $p_c \approx \pi/L^2$, while the probability that the agent does not interact with any other agent is $(1-p_c)^N \approx e^{-\bar{k}}$. This means that the expected number of interacting agents per slot is $\bar{n} \approx N (1-e^{-\bar{k}})$. Solving for $L$ we get
\begin{align}
\label{eq:L}
L \approx \sqrt{-\frac{N\pi}{\ln{(1-\bar{n}/N)}}}.
\end{align}
\item Parameter $\mu_1 > 0$:  We set the initial value of this parameter to $\mu_1 = 0.5$. 
\item Parameters $F_0 \geq 0$, $\mu_2 > 0$: From our experiments we observed that $0.1 \leq F_0 \leq 0.2$ with $\mu_2 = 0.8$ is a good initial configuration for these parameters.
\end{itemize}

\subsection{Parameter tuning}

We next tune the above parameters as described below in order to match the following quantities between simulated and real networks: (i) the average contact duration; (ii) the average number of recurrent components per $10$ minute interval, while ensuring a similar size of the largest component formed; and (iii) the average agent degree in the time-aggregated network. We choose a $10$ minute interval in (ii) in order to give some time to components to break apart (components appearing more than once in an interval are counted only once),  but no more than $10$ minutes to avoid losing the resolution of the components formation. While tuning a specific parameter all other parameters remain fixed, and we take the average of the corresponding metric over $10$ simulation runs.

\begin{figure}[!h]
\centering
\includegraphics[width=17cm,height=6.6cm]{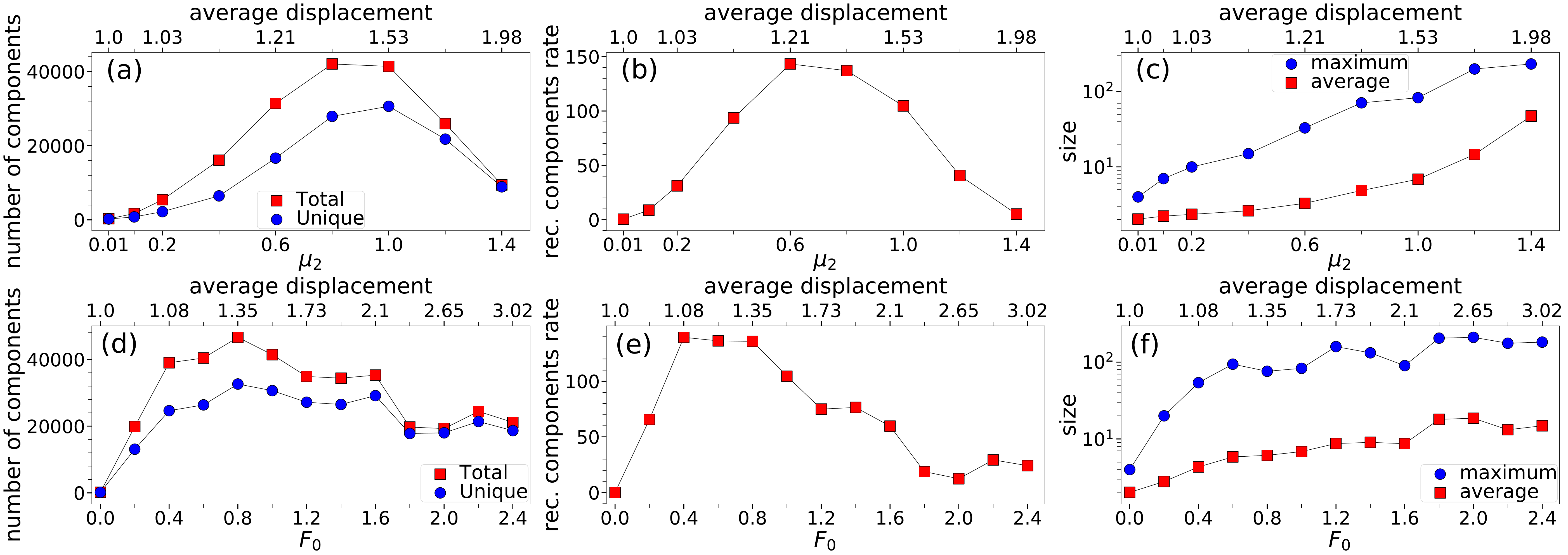}
\caption{Formation and size of components as a function of $\mu_2$ (top row) and $F_0$ (bottom row). In all cases $v=d=1$ and the top $x$-axis indicates the corresponding average agent displacement. \textbf{(a,~d)}~Number of components of at least three agents. The squares show the number of all components formed (unique and recurrent), while the circles show the number of unique components.
The recurrent components are measured within intervals of 10 minutes as described in the text. Each time slot is assumed to be 20 seconds as in the real face-to-face interaction networks, and thus the interval of 10 minutes corresponds to 30 time slots in the simulation. The recurrent components are also measured this way in Fig.~\ref{FigComponents} of the main text. \textbf{(b,~e)}~Corresponding recurrent components rate, i.e., the average number of recurrent components observed in an interval of 10 minutes. \textbf{(c,~f)} Maximum and average size across all components formed (including components of size $2$). Other simulation parameters are $N=242, T=3100, T_{\textnormal{warmup}}=2000,  L=98, \mu_1=0.35$ and the activation probability $r_i$ for each agent $i$ is sampled uniformly at random from $[0, 1]$; these parameter values are also used in Fig.~\ref{FigComponents} of the main text. In (a-c) $F_0=1$, while in (d-f) $\mu_2=1$.
\label{figdisplacement}}
\end{figure}

\begin{enumerate}
\item We tune $\mu_1$ such that the average contact duration in simulations is approximately the same as in the real dataset. (The average contact duration increases with $\mu_1$.)
\item We tune $F_0$ and $\mu_2$ such that the average number of recurrent components per interval of $10$ minutes is approximately the same as in the real dataset, while the size of the largest component formed is similar as in the dataset. (As mentioned in the main text,  as $\mu_2$ increases larger components form, until the agents eventually collapse into a giant connected component. At the same time, the number of components initially increases and then decreases, see Fig.~\ref{FigComponents}(a) in the main text and Figs.~\ref{figdisplacement}a-c below. A similar behavior is observed as $F_0$ increases because the magnitude of the deterministic motion increases compared to the magnitude of the random motion, see Fig.~\ref{FigComponents}(b) in the main text and Figs.~\ref{figdisplacement}d-f below. In this case, an eventual collapse into a giant component can occur if $\mu_2$ is not sufficiently small (Fig.~\ref{FigComponents}(b) in the main text and Figs.~\ref{figdisplacement}d-f). In general, to avoid collapses, as one of these parameters increases the other should decrease.)

\item We tune $L$ such that the average agent degree in the time-aggregated network is approximately the same as in the real network. (Larger values of $L$ result in a smaller average agent degree.)
\item We repeat steps (1)-(3) if needed until all considered metrics ((i)-(iii) above) are approximately the same as in the real dataset.
\end{enumerate}

Finally, as explained in the main text, each agent $i$ is also assigned an activity value $r_i$, which is the probability of the agent to become active at the beginning of each slot if the agent is inactive. In the simulated counterparts of the Hospital and Conference the $r_i$s are sampled uniformly at random from $[0, 1]$. In the simulated counterparts of the Primary and High School we assign $r_i = 0.5$ for all agents $i$, as we have observed that the number of interactions per agent in the corresponding real datasets is somewhat more homogeneous than in the Hospital and Conference datasets. We also assign $r_i=0.5$ for every agent $i$ in the simulations of the MIT Social Evolution.  We note that after tuning the model parameters as described, the resulting average number of interacting agents and links per slot are also similar as in the real networks ($\bar{n}, \bar{l}$ in Table~\ref{tableTraces} of the main text).

In the attractiveness model~\cite{Starnini2013} we also have $v=d=1$ and the free parameters are the number of agents $N$, the number of time slots $T$, and the size of the space $L$, while the $r_i$s are sampled uniformly at random from $[0, 1]$. In our simulations with this model $N$ and $T$ are equal to their counterparts in the real networks, while $L$ is set such that the average number of interacting agents per slot is approximately the same as in the real networks. Specifically, the values of $L$  for the simulated networks of the Hospital, Primary School, High School, Conference and MIT Social Evolution are $L= 44, 50, 80, 85, 45$, respectively. A warmup period is not required.

\section{Agent displacement}
\label{sec:displacement}

In this section we analyze the expected agent displacement in the FDM, $E[\Delta r]$. As in the main text, let $\mathcal{S}(t)$ be the set of moving and interacting agents in slot $t$ and $F_i^{t,x}, F_i^{t,y}$ the total attractive forces exerted to a moving agent $i$ by all agents $j \in \mathcal{S}(t)$ along the $x$ and $y$ directions of the motion,
\begin{align}
\label{eq:not_1}
F_i^{t,x} & = \sum_{j \in \mathcal{S}(t)}F_{ij}\cos{\psi_{ij}^t},\textnormal{~}\cos{\psi_{ij}^{t}} = \frac{(x^t_j - x^t_i)}{\sqrt{(x^t_j - x^t_i)^2+(y^t_j - y^t_i)^2}},\\
\label{eq:not_2}
F_i^{t,y} &= \sum_{j\in \mathcal{S}(t)} F_{ij}\sin{\psi_{ij}^t}, \textnormal{~}\sin{\psi_{ij}^{t}} = \frac{(y^t_j - y^t_i)}{\sqrt{(x^t_j - x^t_i)^2+(y^t_j - y^t_i)^2}},
\end{align}
where $F_{ij}=F_0 e^{-\frac{s_{ij}}{\mu_2}}$ and $s_{ij} = N(\pi - \vert\pi-\vert\theta_i - \theta_j\vert\vert)/2\pi$ is the similarity distance between $i$ and $j$. Since the similarity coordinates are uniformly distributed we can set without loss of generality $\theta_i=0$, and compute the second moment of $F_{ij}$,
\begin{equation}
\label{eq:force_moment}
E[F_{ij}^2]= \frac{1}{2 \pi} \int_{0}^{2\pi} F_{ij}^2 \mathrm{d}\theta_j=\frac{F_0^2 \mu_2}{N}(1-e^{-\frac{N}{\mu_2}}) \approx \frac{F_0^2 \mu_2}{N},
\end{equation}
where the last approximation holds for large $N/\mu_2$. Further, assuming that $\psi_{ij}^t$ in Eqs.~(\ref{eq:not_1}),~(\ref{eq:not_2}) is uniformly distributed on $[0, 2\pi]$, i.e., assuming that the agents $j\in \mathcal{S}(t)$ are uniformly distributed around agent $i$ in the Euclidean space, we have $E[\cos{\psi_{ij}^t}]=E[\sin{\psi_{ij}^t}]=0$, and
\begin{equation}
\label{eq:psi_moment}
E[(\cos{\psi_{ij}^t})^2]=E[(\sin{\psi_{ij}^t})^2]=\frac{1}{2\pi} \int_{0}^{2\pi} (\sin{\psi_{ij}^t})^2 \mathrm{d}\psi_{ij}^t=\frac{1}{2}.
\end{equation}
Using Eqs.~(\ref{eq:not_1})-(\ref{eq:psi_moment}) we can write
\begin{equation}
\label{eq:not_3}
E[(F_i^{t,x})^2 | \mathcal{S}(t)]=E[(F_i^{t,y})^2 | \mathcal{S}(t)]=E\left[\left(\sum_{j \in \mathcal{S}(t)}F_{ij}\sin{\psi_{ij}^t}\right)^2\right] = \sum_{j \in \mathcal{S}(t)} E[F_{ij}^2] E[(\sin{\psi_{ij}^t})^2] \approx \frac{F_0^2 \mu_2}{2} \frac{|\mathcal{S}(t)|}{N}.
\end{equation}
The above relation depends only on the number of moving and interacting agents, $|\mathcal{S}(t)|$, and not on the exact agent $i$ or the agents $j \in \mathcal{S}(t)$. Furthermore, the average number of moving and interacting agents per slot is $\bar{r}N+(1-\bar{r})\bar{n}$, where $\bar{r}$ is the average agent activation probability~\footnote{\label{noteActAgents} Given that the expected number of interacting agents per slot is $\bar{n}$ and that we activate each of the $N-\bar{n}$ inactive agents with an average probability $\bar{r}$, the expected number of moving and interacting agents per slot is $\bar{n}+(N - \bar{n})\bar{r}=\bar{r}N+(1-\bar{r})\bar{n}$.}. Therefore, removing the condition on the index $i$ and slot $t$ we can write
\begin{equation}
E[(F^x)^2]=E[(F^y)^2] \approx \frac{F_0^2 \mu_2}{2} \left(\bar{r}+(1-\bar{r})\frac{\bar{n}}{N}\right).
\end{equation}
Now, from Eqs.~(\ref{eq:motion_eqs_1}),~(\ref{eq:motion_eqs_2}) in the main text, the expected displacement of an agent $i$ in slot $t$, $E[\Delta r_i^t|\mathcal{S}(t)]$, is
\begin{align}
\label{eq:upper_bound}
\nonumber E[\Delta r_i^t | \mathcal{S}(t)]&=E\left[\sqrt{(x_i^{t+1}-x_i^{t})^2+(y_i^{t+1}-y_i^{t})^2}| \mathcal{S}(t)\right]=E\left[\sqrt{\left(F_i^{t,x}+ R^x_i\right)^2+\left(F_i^{t,y}  + R^y_i\right)^2}|\mathcal{S}(t)\right]\\
\nonumber &\leq \sqrt{E\left[\left(F_i^{t,x} + R^x_i\right)^2|\mathcal{S}(t)\right]+E\left[\left(F_i^{t,y} + R^y_i\right)^2|\mathcal{S}(t)\right]}= \sqrt{2 E[(F_i^{t,x})^2|\mathcal{S}(t)]+ 2 E[(R^x_i)^2]}\\
& = \sqrt{F_0^2 \mu_2 \frac{|S(t)|}{N}+ v^2}.
\end{align}
As above, we can remove the condition on the index $i$ and slot $t$ and write
\begin{equation}
\label{eq:upper_bound_uncon}
E[\Delta r] \leq  \sqrt{F_0^2 \mu_2 \left(\bar{r}+(1-\bar{r})\frac{\bar{n}}{N}\right)+ v^2}.
\end{equation}
The inequalities in Eqs.~(\ref{eq:upper_bound}),~(\ref{eq:upper_bound_uncon}) hold since $E[\sqrt{x}] \leq \sqrt{E[x]}$ for $x  \geq 0$ (Jensen's inequality for concave functions). Further, since $R^x_i= v\cos\phi_i$, $R^y_i = v\sin\phi_i$, where $\phi_i$ is sampled uniformly at random from $[0, 2\pi$] (see main text), we also use in Eq.~(\ref{eq:upper_bound}) the facts $E[R^x_i]=E[R^y_i]=0$ and $E[(R^x_i)^2]=E[(R^y_i)^2]=v^2/2$. Finally, we note that the magnitude of the random displacement is always $\sqrt{(R^x_i)^2+(R^y_i)^2}=v$. 

Table~\ref{tableParameters} shows the values of $F_0, \mu_2, N$, while Table~\ref{tableTraces} in the main text shows the values of $\bar{n}$. Parameter $v$ is fixed to $v=d=1$ and in all of our simulated networks $\bar{r}=1/2$ (Sec. \ref{sec:parameter_tuning}). The corresponding average displacement per slot in the simulations of the Hospital, Primary School, High School, Conference and MIT Social Evolution is $1.00356, 1.0090, 1.0023, 1.0005, 1.00271$ while the corresponding upper bounds predicted by Eq.~(\ref{eq:upper_bound_uncon}) are $1.00359, 1.0096, 1.0029, 1.0004, 1.00278$. We can see that the random displacement is significantly larger than the displacement due to the attraction forces. Specifically,  with the Hospital, Primary School, High School, Conference and MIT Social Evolution parameters we have $F_0\sqrt{\mu_2\left(\bar{r}+(1-\bar{r})\frac{\bar{n}}{N}\right)}=0.084, 0.139, 0.077, 0.028, 0.075$ vs. $v=1$. We note that in general a natural choice for the total expected displacement is to be in the order of the interaction range $d$. This will give the chance to escaping agents to move away from their interactions in one time slot, without drifting far away.

\section{Similarity forces in spaces with broken triangle inequality}
\label{sec:metric_space}

The key metric property of the similarity space, i.e., the triangle inequality, ensures that if an agent $a$ is close to an agent $b$ and $b$ is close to a third agent $c$, then $c$ is also close to $a$. This means that the forces between all the three agents are strong and these agents will tend to gather close to each other in the Euclidean space forming triangle $abc$. In other words, the triangle inequality in the similarity space imposes a localization effect on the forces, which attract similar agents to form clusters in the observed network. If the forces decrease fast enough with the similarity distance, then we indeed expect to see an abundance of small connected components as in the real datasets and the model (Figs.~\ref{figHPProps}f-\ref{figMITProps}f). On the other hand, if the similarity distances do not satisfy the triangle inequality, then agents $a$ and $c$ might not be close to each other, but instead close to some other agents $d$ and $e$, forming chain $dabce$ in the observed network. That is, if the similarity space does not have a metric structure, forces loose their localization, and agents tend to form larger components.

To verify these arguments we break the triangle inequality in the similarity space by assigning similarity distances sampled uniformly from $[0, \pi R]$  to all pairs of agents (non-metric case), instead of assigning to the agents similarity coordinates on the circle (metric case). We see in Figs.~\ref{figHPProps}f-\ref{figMITProps}f that indeed in the non-metric case larger components form even though the values of the simulation parameters are set exactly as in the metric case. 

\begin{figure}[!h]
\centering
\includegraphics[width=17cm,height=9cm]{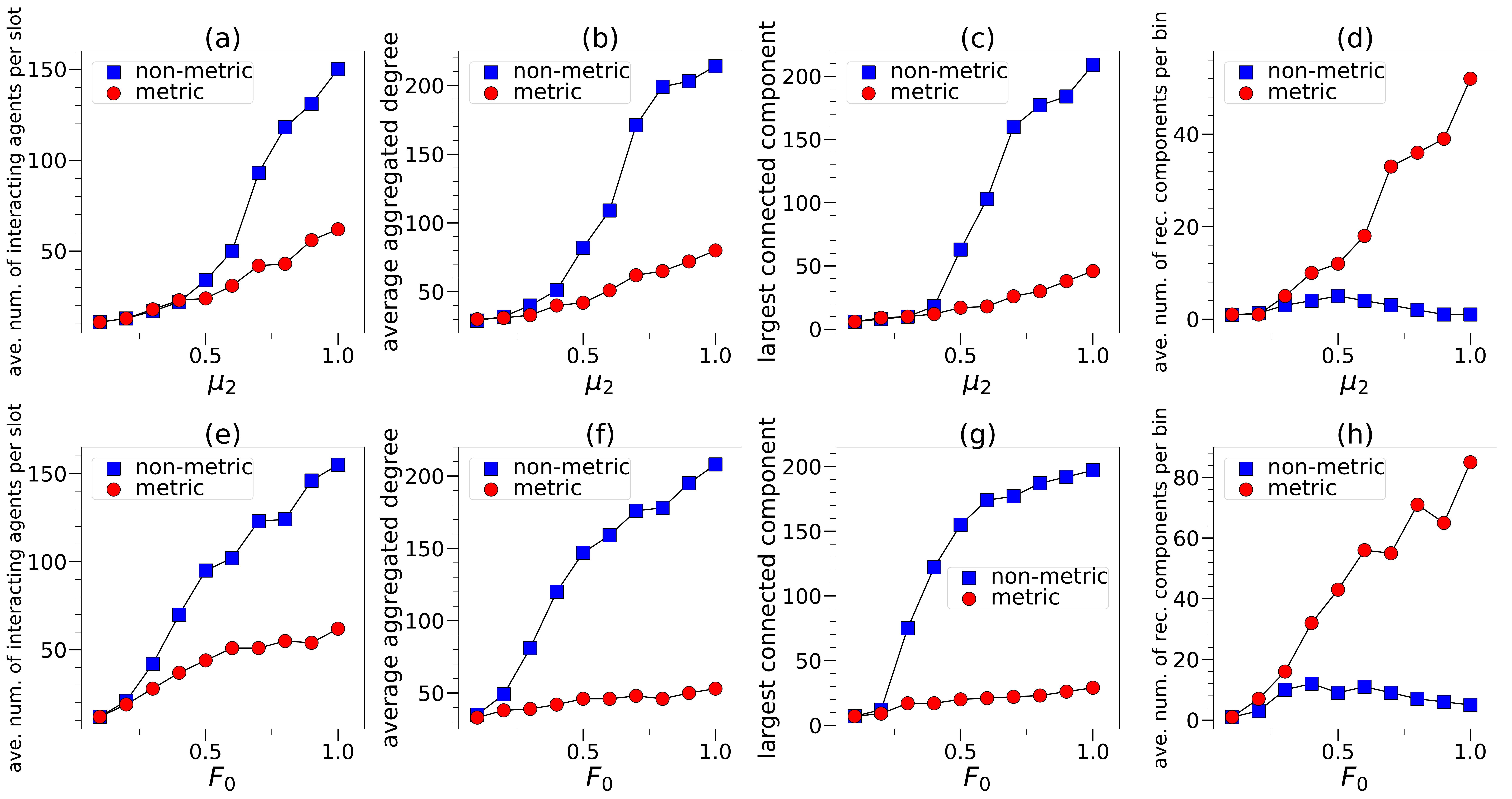}
\caption{Similarity spaces with metric vs. non-metric structure.
\textbf{(a,~e)}~Average number of interacting agents per slot in FDM simulated networks with and without metric structure in the similarity space, as a function of $\mu_2$ and $F_0$. In (a) $F_0 = 0.2$ and in (e) $\mu_2=0.4$.
\textbf{(b,~f)}~Average agent degree in the time-aggregated network of contacts for the networks in (a, e).
\textbf{(c,~g)}~Size of the largest component formed in the networks of (a, e).
\textbf{(d,~h)}~Average number of recurrent components per interval (bin) of $10$ minutes in the networks of (a, e).
Each point in the plots is an average over $10$ simulation runs.
\label{figGeomStats}}
\end{figure}

\begin{figure}[!h]
\centering
\includegraphics[width=10cm,height=4cm]{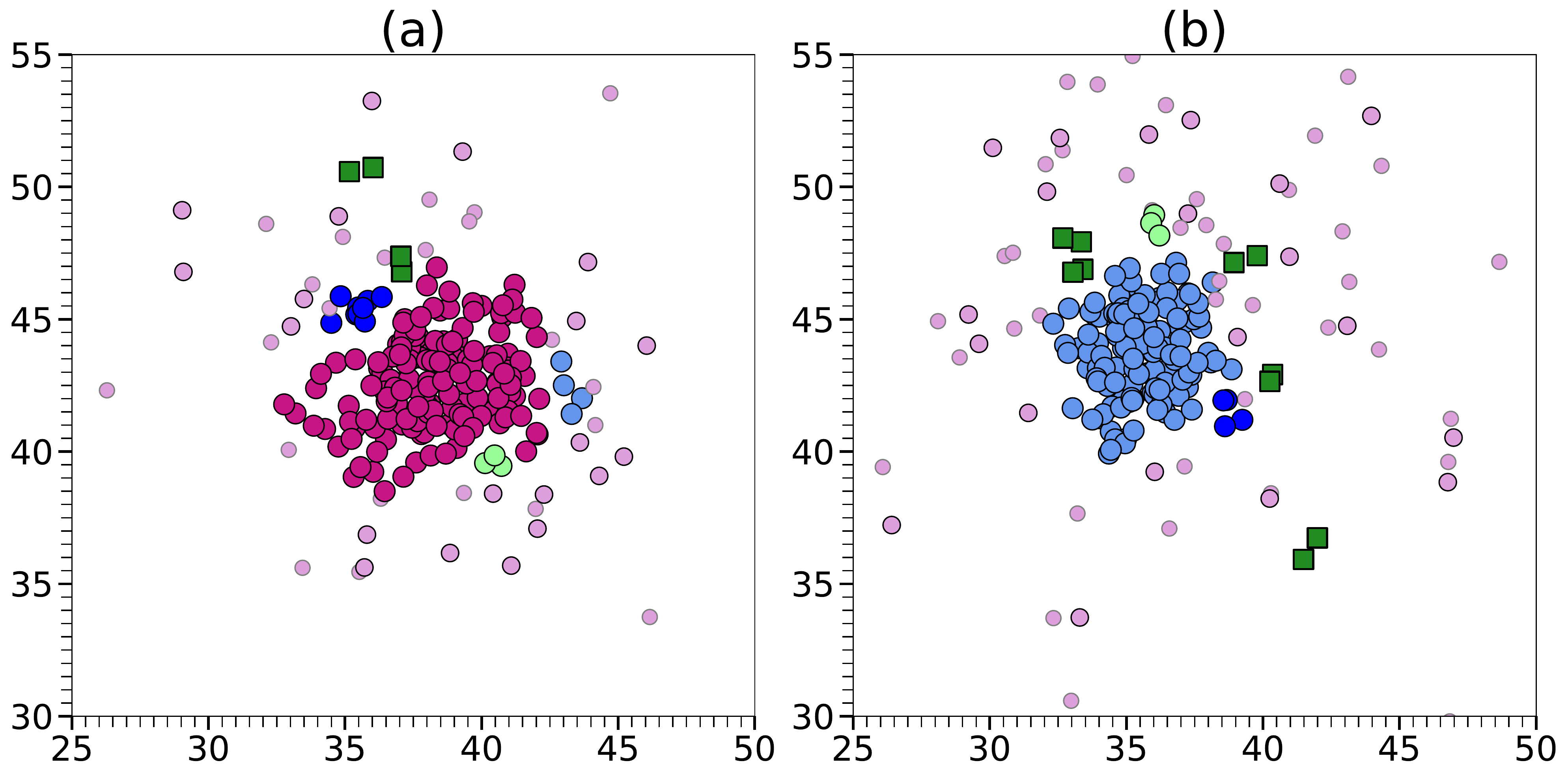}
\caption{Snapshots of collapsing agents in the middle of the Euclidean space. \textbf{(a)} Snapshot of the agents in the Euclidean space at time slot $t = 2677$ (total slots $= 3100$) in a simulated network of Fig.~\ref{figGeomStats} with non-metric similarity space and $\mu_2=0.8, F_0=0.2$. \textbf{(b)} Same as (a) but for $F_0=0.7, \mu_2=0.4$ at slot $t=2981$. In both (a, b) the smallest grayed-out circles are inactive agents, the second smallest circles are moving agents that are not interacting, the dark green squares are connected components of $2$ agents, and the largest multicolored circles are connected components of least $3$ agents. In (a) the agents in the giant connected component are colored purple, while in (b) they are colored light blue.
 \label{figGeomSnaps}}
\end{figure}

Furthermore, in Fig.~\ref{figGeomStats} we consider simulation runs with the FDM and the Primary School parameters in Table~\ref{tableParameters}, except that in Figs.~\ref{figGeomStats}a-d we gradually increase $\mu_2$ from $0.1$ to $1$, while in Figs.~\ref{figGeomStats}e-h we gradually increase $F_0$ from $0.1$ to $1$ with $\mu_2 = 0.4$. We see that in the non-metric case, as $\mu_2$ or $F_0$ increases, the average number of interacting agents per slot increases much faster than in the metric case (Figs.~\ref{figGeomStats}a,e). This is also the case for the average agent degree in the time-aggregated network (Figs.~\ref{figGeomStats}b,f) and the size of the largest component (Figs.~\ref{figGeomStats}c,g). We also see that in the non-metric case neither $\mu_2$ nor $F_0$ can increase the average number of recurrent components per interval of $10$ minutes beyond a certain value (Figs.~\ref{figGeomStats}d,h), as most agents collapse into a giant connected component. Specifically, in the non-metric case agents start forming significantly larger and larger components after $\mu_2 = 0.5$ and $F_0 = 0.2$ (Figs~\ref{figGeomStats}c,g). For $\mu_2 \geq 0.7$ and $F_0 \geq 0.4$ the agents collapse into a giant component in the middle of the Euclidean space (cf. Fig.~\ref{figGeomSnaps}) and remain collapsed until the end of the simulation. By contrast, in the metric case the corresponding increases in Figs.~\ref{figGeomStats}a-c,e-g are much more gradual and agents do not collapse into a giant component. Furthermore, the average number of recurrent components per interval of $10$ minutes increases with $\mu_2$ and $F_0$ (Figs.~\ref{figGeomStats}d,h). Therefore, the metric structure of the similarity space promotes the formation of sparse network snapshots without giant connected components, as in the real networks.

\section{SIS spreading in real and modeled networks}
\label{sec:sis}

Here we provide more details on the susceptible-infected-susceptible (SIS) epidemic spreading model~\cite{sis_ref} considered in the main text. As mentioned in the main text, in the SIS model each agent can be in one of two states at any time slot $t$, susceptible~(S) or infected~(I). At any time slot an infected agent recovers with probability $\beta$ and becomes susceptible again, whereas infected agents infect the susceptible agents with whom they interact, with probability $\alpha$. Therefore, the only transition of states is $\textnormal{S}\rightarrow \textnormal{I} \rightarrow \textnormal{S}$. 

To obtain the results in Fig.~\ref{FigSpread} of the main text we have used the dSIS (dynamic SIS) model for temporal networks from the Network Diffusion Library~\cite{NDlib}. For each simulation of the process we compute the percentage of infected agents per slot, and then take the average of this percentage over the considered slots (prevalence). We consider the first activity cycle of the Hospital and Primary school and the second cycle of the High School---we consider the second cycle of the High School as its first cycle has fewer recorded slots than the rest of its cycles (Sec.~\ref{sec:datasets}). In all cases the results are similar in all activity cycles of similar durations. The corresponding simulated networks with the FDM (Table~\ref{tableParameters}) are run (after $T_{\textnormal{warmup}}$) for the same duration as the corresponding cycles in the real networks and the prevalence is measured excluding the $T_{\textnormal{warmup}}$ period. In the real networks the results are averages over $20$ simulated SIS processes. The results with the FDM are averages across $10$ simulated counterparts of each real network; in each counterpart the prevalence is averaged over $5$ SIS processes. In all cases each SIS process has a different initial set of infected agents that consists of $10\%$ of all agents selected at random. 

\begin{figure}[!h]
\includegraphics[width=7cm]{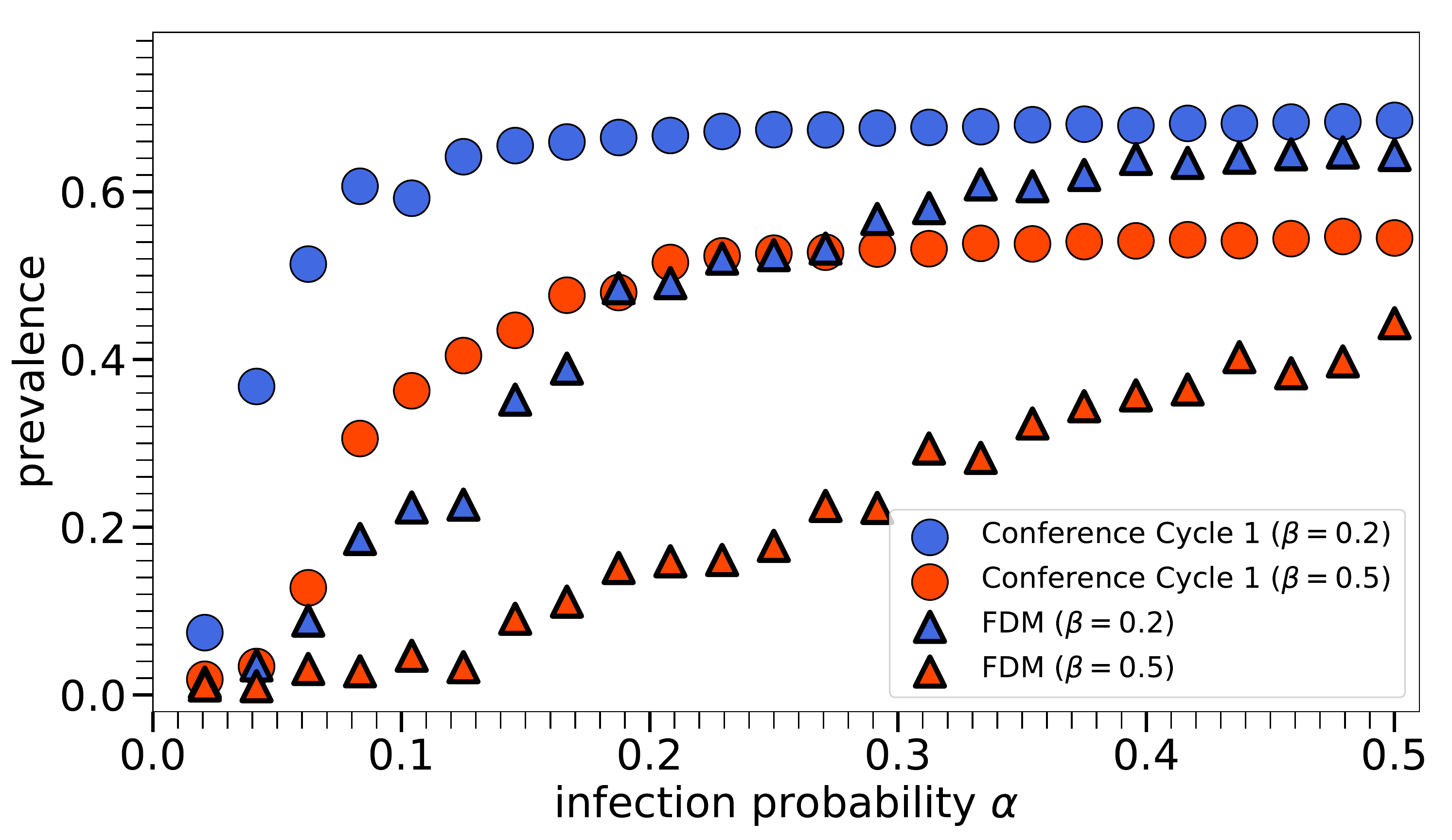}
\caption{Same as Fig.~\ref{FigSpread} in the main text but for the Conference dataset.
\label{figSISConf}}
\end{figure}

In Fig.~\ref{figSISConf} we also report prevalence results for the Conference (first activity cycle). As can be seen, in this case the SIS process performs differently than in the corresponding FDM networks. This was expected since as explained in Sec.~\ref{SecProps} this dataset has some different properties from the rest of the datasets we consider (cf. Fig.~\ref{figHYTProps}h and the related discussion in Sec.~\ref{SecProps}). 

\section{Non-uniform similarity coordinates}
\label{sec:non_uniform}

In this section we consider a non-uniform distribution of the similarity coordinates corresponding to the organization of agents into communities. To this end, we sample the angular coordinates of nodes from a Gaussian mixture distribution (GMD) as in~\cite{CannistracinPSO}. The GMD is a mixture of multiple Gaussian distribution components, characterized by the following parameters~\cite{gaussian_mixture, CannistracinPSO}: (i) $C>0$, which is the number of components, each one representative of a community; (ii)  $\mu_{1...C} \in [0,2\pi]$, which are the means of every component, representing the central locations of the communities in the angular space; (iii) $\sigma_{1...C} > 0$, which are the standard deviations of every component, determining how much the communities are spread in the angular space; and (iv) $\rho_{1...C}$ $(\sum_i \rho_i = 1)$, which are the mixing proportions of every component, determining the relative sizes of the communities.

\begin{figure}[!h]
\includegraphics[width=15cm,height=6cm]{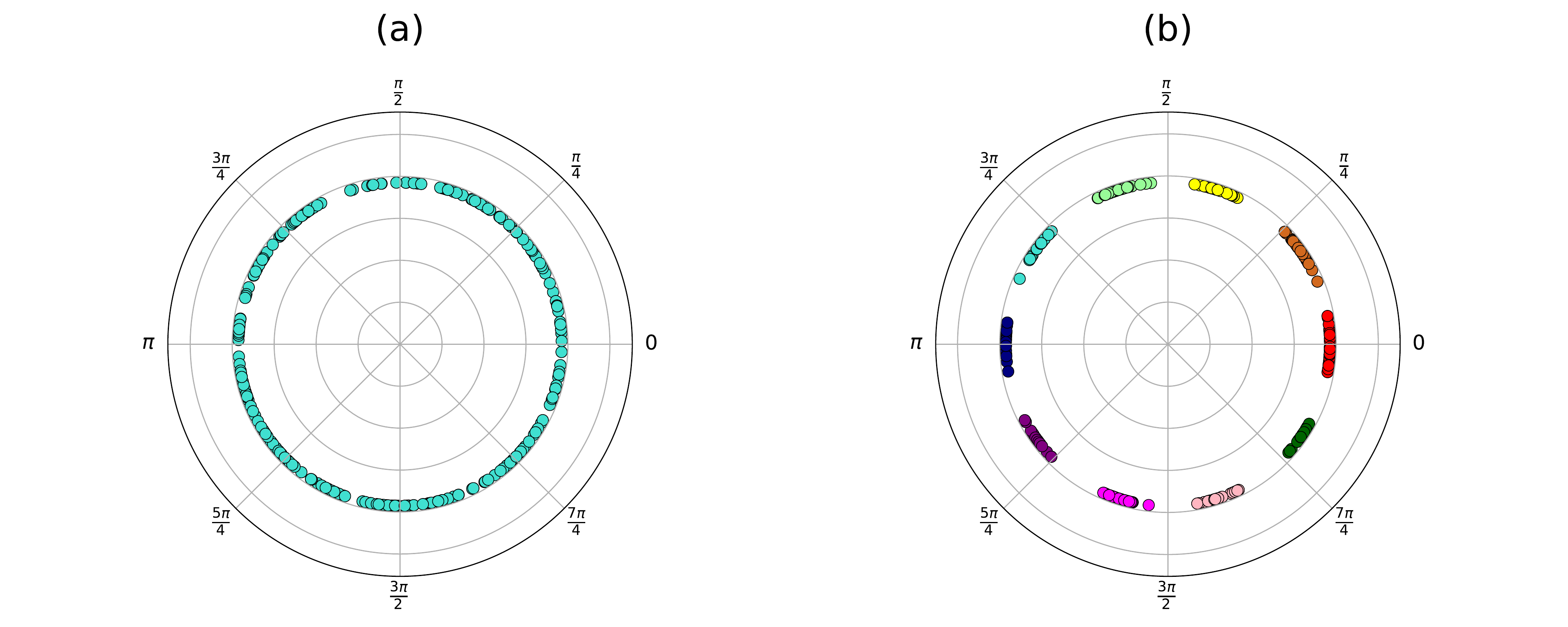}
\caption{\textbf{(a)} Uniform distribution of the similarity coordinates. \textbf{(b)} Non-uniform distribution of the similarity coordinates corresponding to the separation of agents into $10$ communities, each indicated by a different color.
\label{figPS_S1Spaces}}
\end{figure}

We consider simulations of the Primary School. Since in the Primary school students are divided into $10$ classes~\cite{PrimaryData}, we assume that there are $10$ communities and sample the angular coordinates of the agents from a GMD with parameters $C = 10$, $\mu_i= 2\pi (i-1)/C$, $\sigma_i = 2\pi/(8C)$, $\rho_i=1/C$, $i=1 \ldots C$. Fig.~\ref{figPS_S1Spaces}(b) visualizes the distribution of the agents' coordinates and juxtaposes it against the uniform distribution (Fig.~\ref{figPS_S1Spaces}(a)). We can see that the agents are divided into $10$ distinct communities in the similarity space with each community having a similar number of agents. 

\begin{figure}[!h]
\includegraphics[width=17cm]{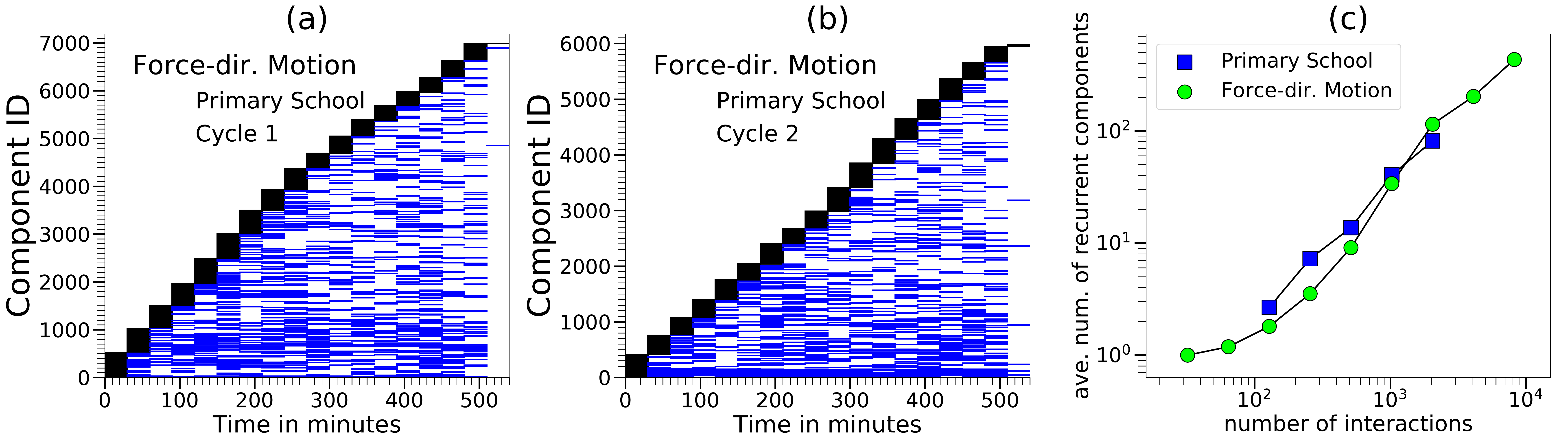}
\caption{\textbf{(a, b)}~Unique and recurrent components found in a simulation run of the FDM (Force-dir.~Motion) model with the non-uniform similarity coordinates in Fig.~\ref{figPS_S1Spaces}(b), assuming activity cycles of the same durations as in the Primary School. \textbf{(c)}~Average number of recurrent components where a node participates as a function of the total number of interactions of the node in the Primary School and in simulated networks with the non-uniform similarity coordinates in Fig.~\ref{figPS_S1Spaces}(b). The result in (c) is an average over $10$ simulation runs. 
\label{figPSComps_NonUnif}}
\end{figure}

\begin{figure}[!h]
\includegraphics[width=17.8cm,height=6.2cm]{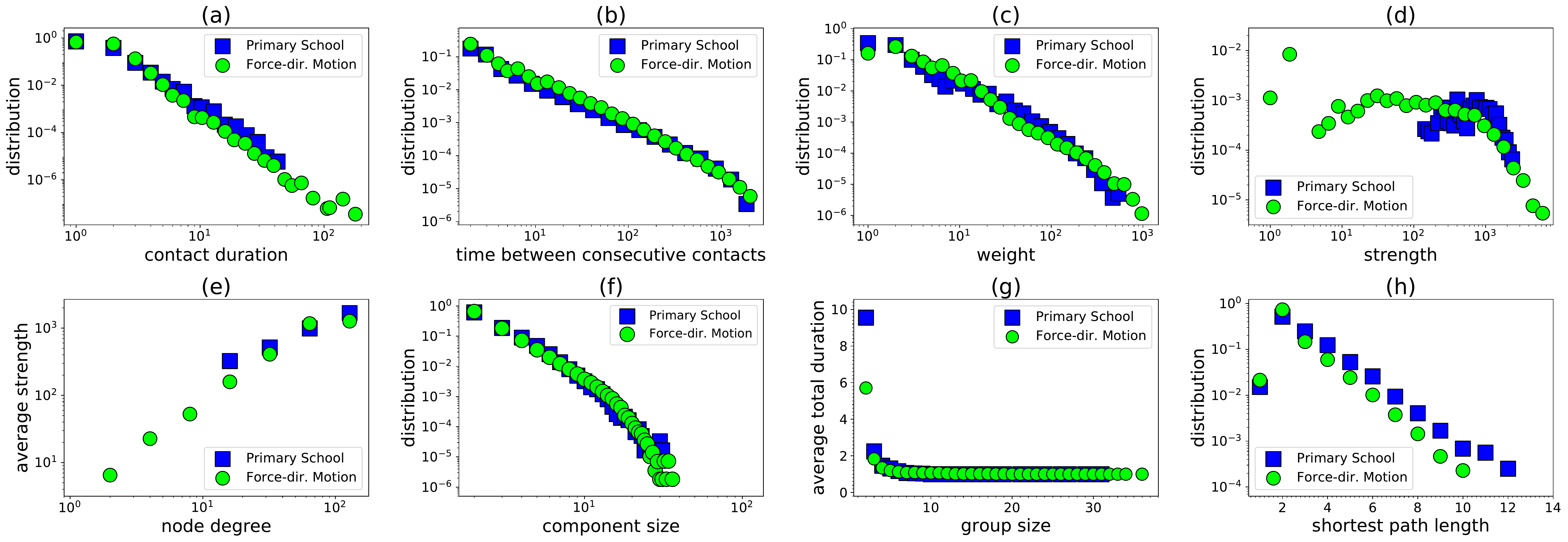}
\caption{Properties of the Primary School face-to-face interaction network and of corresponding simulated networks with the FDM (Force-dir.~Motion) model with the non-uniform similarity coordinates in Fig.~\ref{figPS_S1Spaces}(b). In all cases the simulation results are averages over $10$ runs.
\label{figPSProps_NonUnif}}
\end{figure}

We tune the model parameters $L$, $\mu_1$, $F_0$, $\mu_2$ as described in Sec.~\ref{sec:parameter_tuning}, obtaining: $L = 83$, $\mu_1 = 0.1$, $F_0 = 0.2$, $\mu_2 = 0.32$. The rest of the parameters are as in the simulations of the Primary School in Sec.~\ref{sec:parameter_tuning}. In Figs.~\ref{figPSComps_NonUnif}, \ref{figPSProps_NonUnif} we see that the results are very similar to the ones in Secs.~\ref{sec:recurrent_components},~\ref{SecProps} where the similarity coordinates were uniformly distributed. In other words, the organization of agents into communities does not affect the results.

\section{Hyperbolic space considerations}
\label{sec:hyperbolic}

Finally, as mentioned in the main text, hyperbolic spaces appear as the most natural geometric spaces underlying the observed topologies of traditional complex networks, whose degree distributions are heterogeneous~\cite{Krioukov2010}. In addition to similarity coordinates $\theta$s, nodes in these spaces also have popularity coordinates $r$s, and the hidden distance between two nodes is not  just the angular distance $R\Delta\theta$ but the effective distance $\chi=R\Delta\theta/(\kappa \kappa')$, where $\kappa, \kappa'$ are the expected degrees of the nodes, $\kappa \sim e^{-r}$~\cite{Krioukov2010, Papadopoulos2012}. 

\begin{figure}[!h]
\includegraphics[width=8cm,height=5cm]{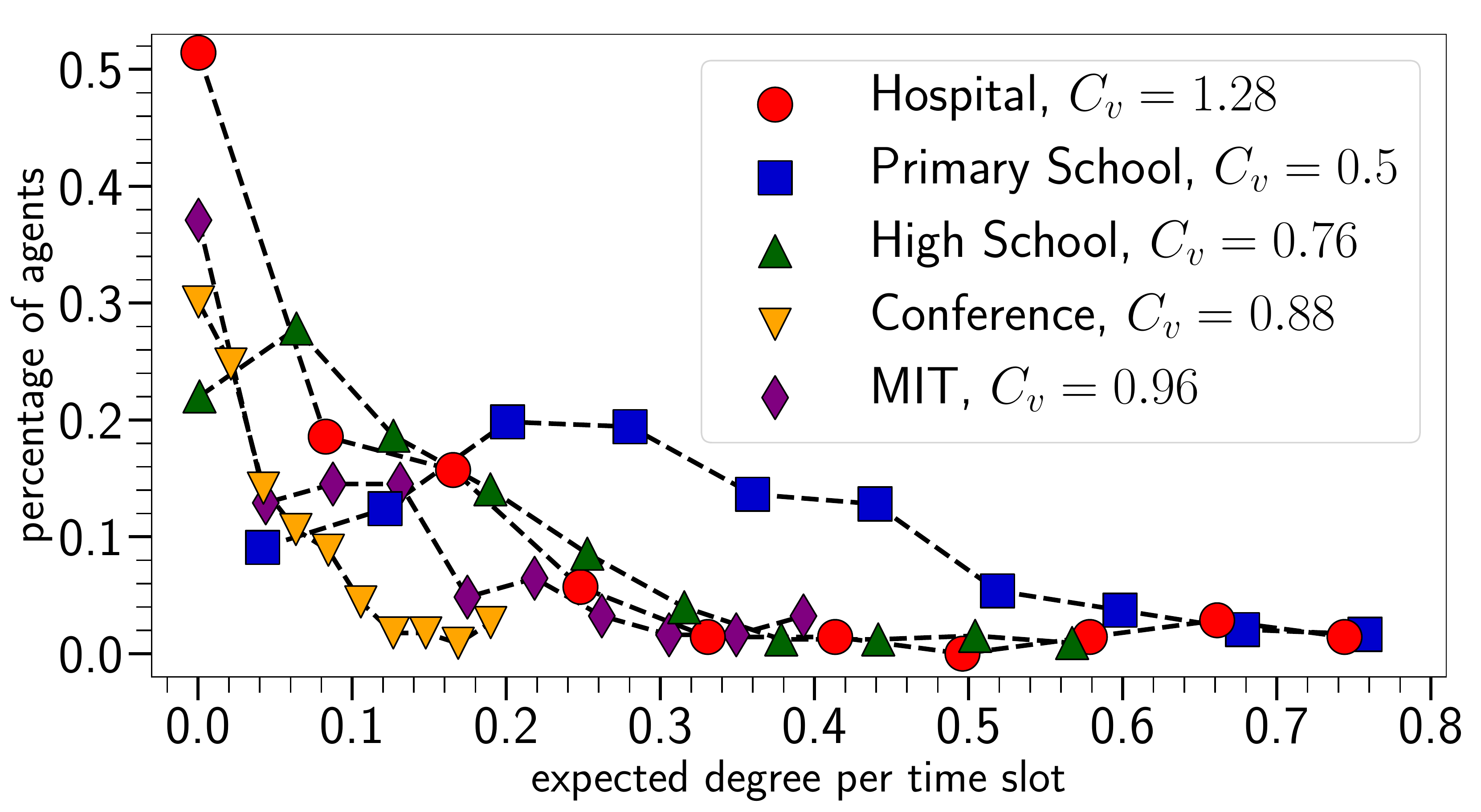}
\caption{Distribution of the expected agent degree per time slot in the real data. The $C_v$s in the legend indicate the coefficient of variation (ratio of the standard deviation to the mean) of each distribution. 
\label{figKappas}}
\end{figure}

\begin{figure}[!h]
\includegraphics[width=18cm, height=3.7cm]{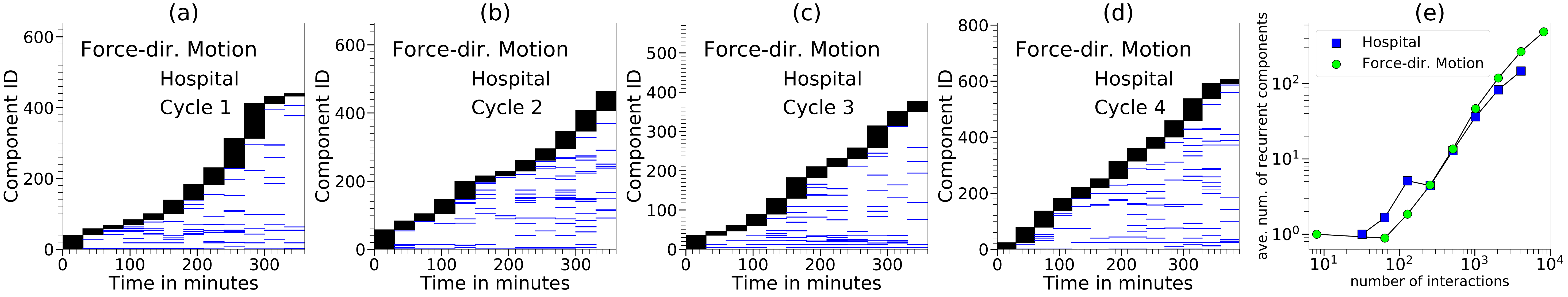}
\caption{\textbf{(a-d)}~Unique and recurrent components found in a simulation run of the FDM (Force-dir.~Motion) model that uses effective distances, in activity cycles of the same durations as in the Hospital. \textbf{(e)} Average number of recurrent components where a node participates as a function of the total number of interactions of the node in the Hospital and in simulated networks with the FDM that uses effective distances. The result in (e) is an average over $10$ simulation runs.
\label{figHPComps_Popularity}}
\end{figure}

\begin{figure}[!h]
\includegraphics[width=14cm, height=4.5cm]{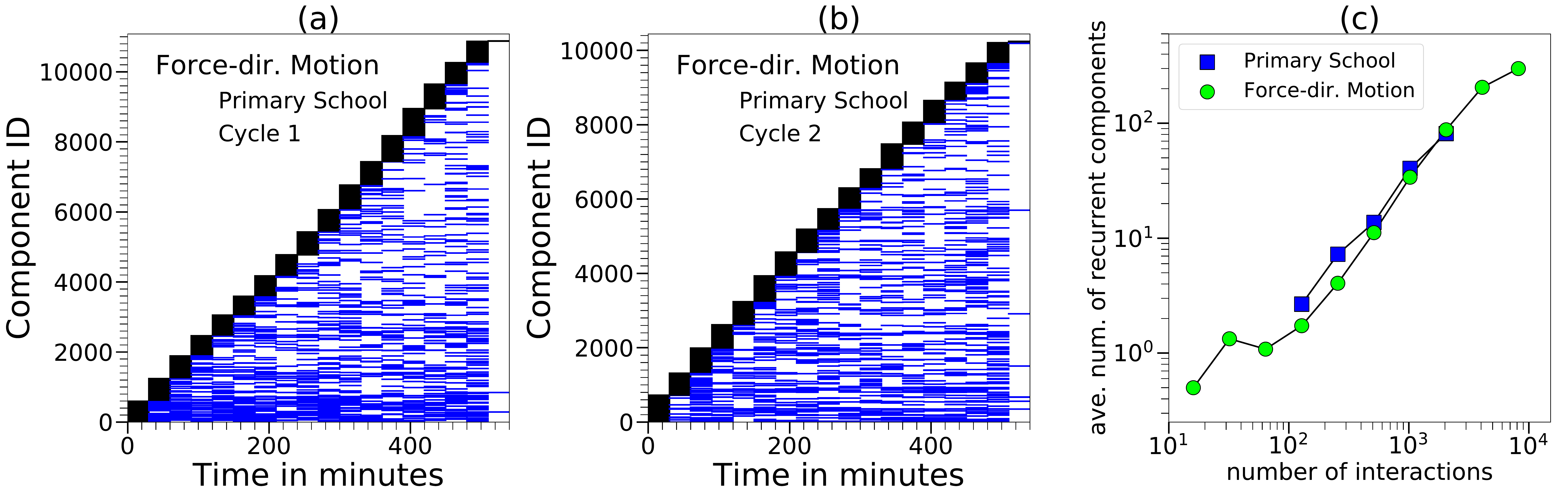}
\caption{Same as Fig.~\ref{figHPComps_Popularity} but for the Primary School.
\label{figPSComps_Popularity}}
\end{figure}

\begin{figure}[!h]
\includegraphics[width=13cm]{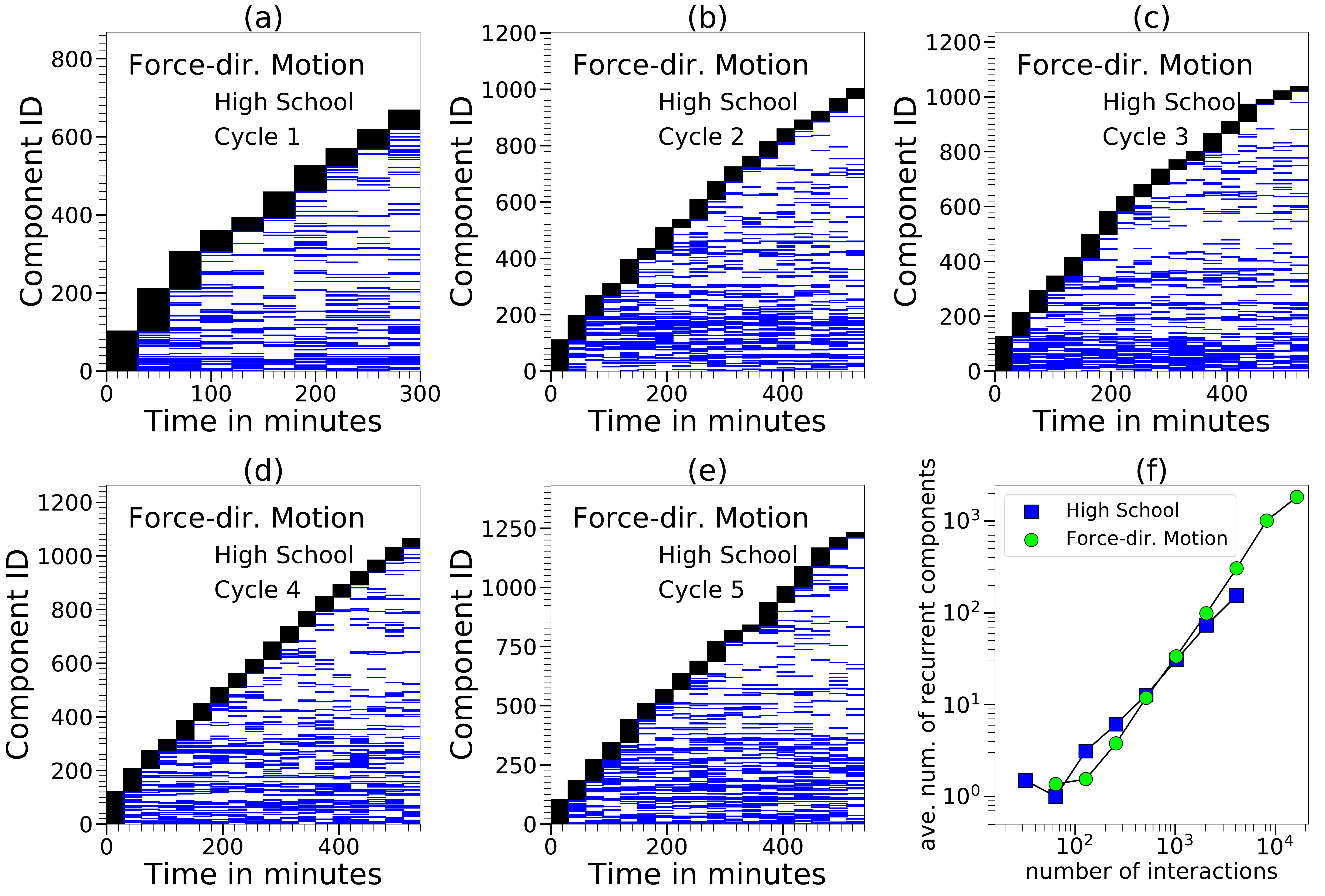}
\caption{Same as Fig.~\ref{figHPComps_Popularity} but for the High School. 
\label{figHSComps_Popularity}}

\end{figure}

\begin{figure}[!h]
\includegraphics[width=10cm]{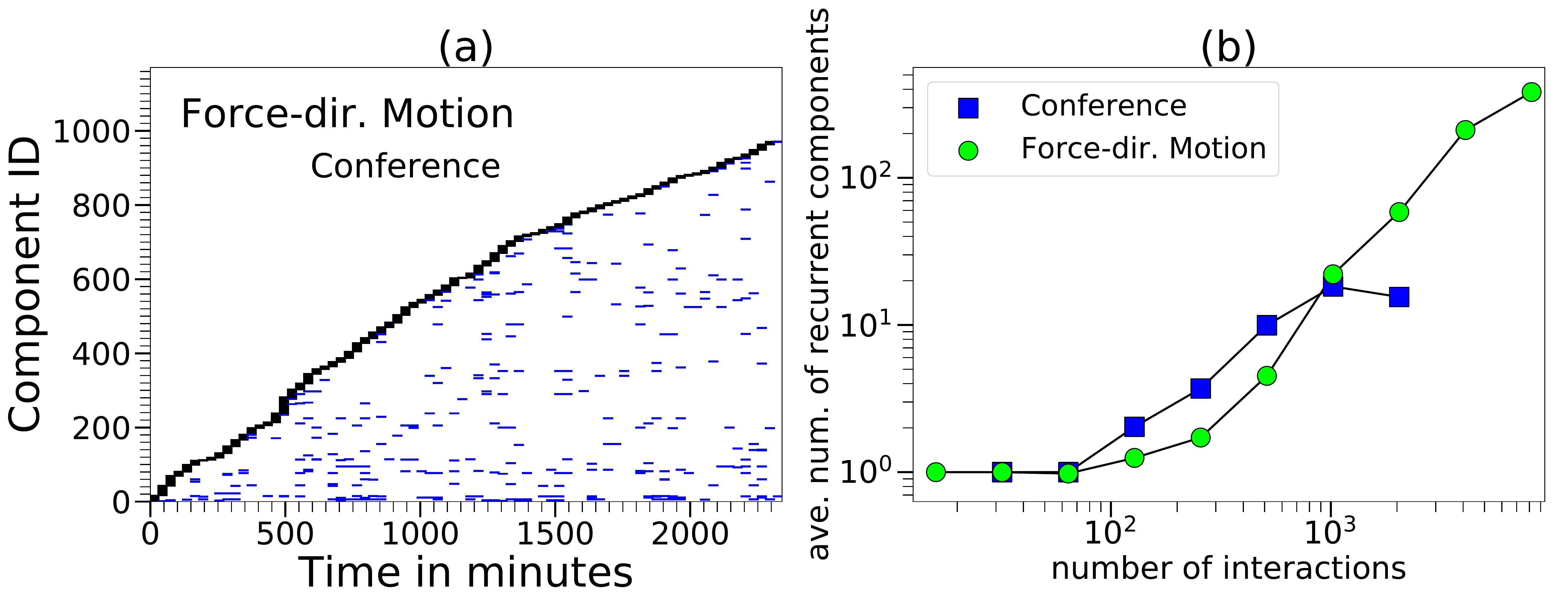}
\caption{Same as Fig.~\ref{figHPComps_Popularity} but for the Conference.
\label{figHYTComps_Popularity}}
\end{figure}

\begin{figure}[!h]
\centering
\includegraphics[width=17.8cm,height=6.2cm]{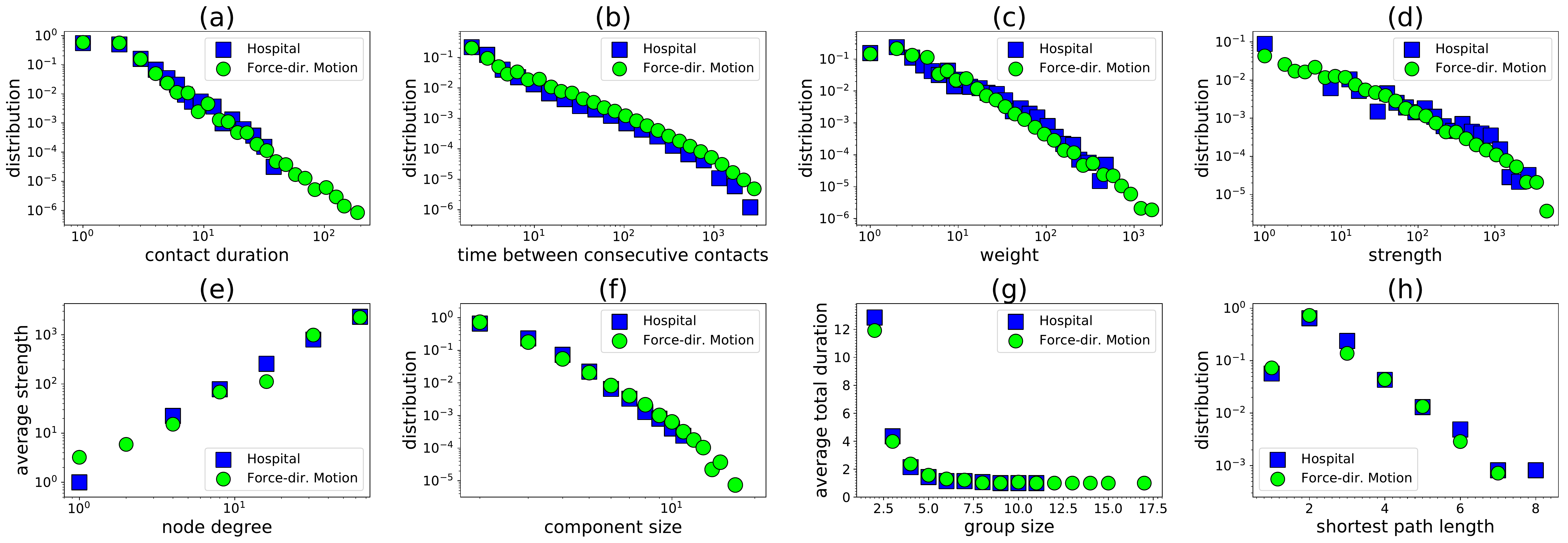}
\caption{Properties of the Hospital face-to-face interaction network and of corresponding simulated networks with the FDM (Force-dir.~Motion) model that uses effective distances. The results are averages over $10$ simulation runs.
\label{figHPProps_Popularity}}
\end{figure}

\begin{figure}[!h]
\centering
\includegraphics[width=17.8cm,height=6.2cm]{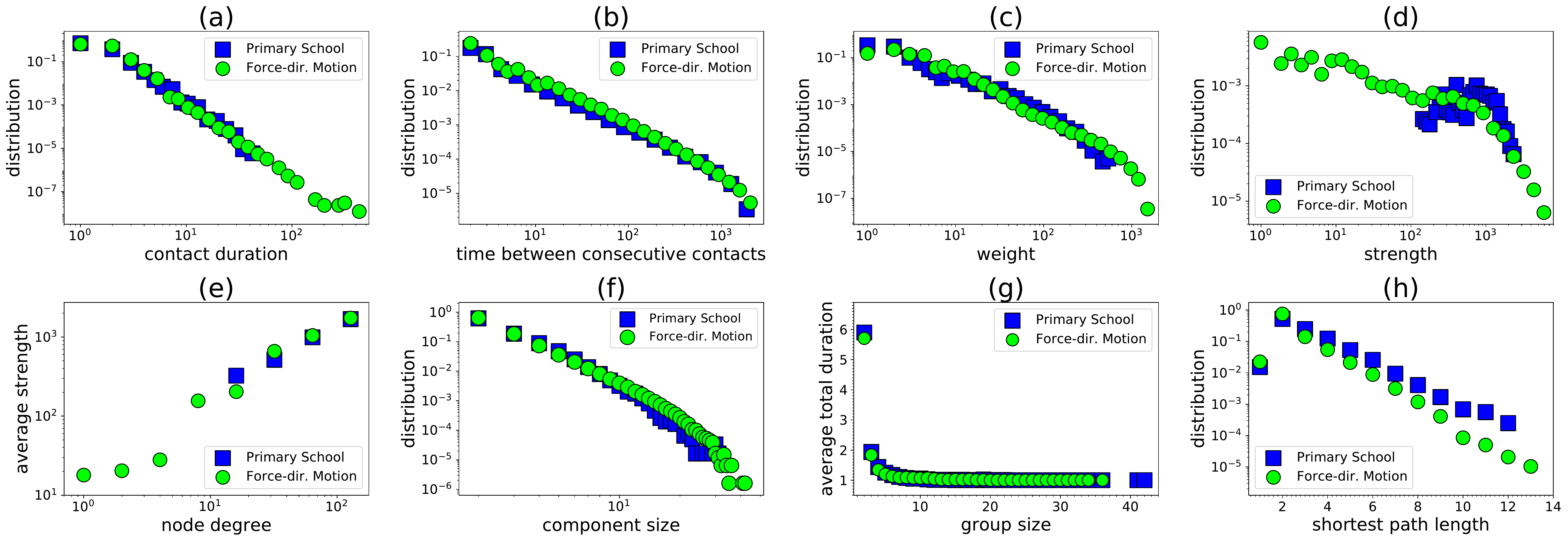}
\caption{Same as Fig.~\ref{figHPProps_Popularity} but for the Primary School.
\label{figPSProps_Popularity}}
\end{figure}

\begin{figure}[!h]
\centering
\includegraphics[width=17.8cm,height=6.2cm]{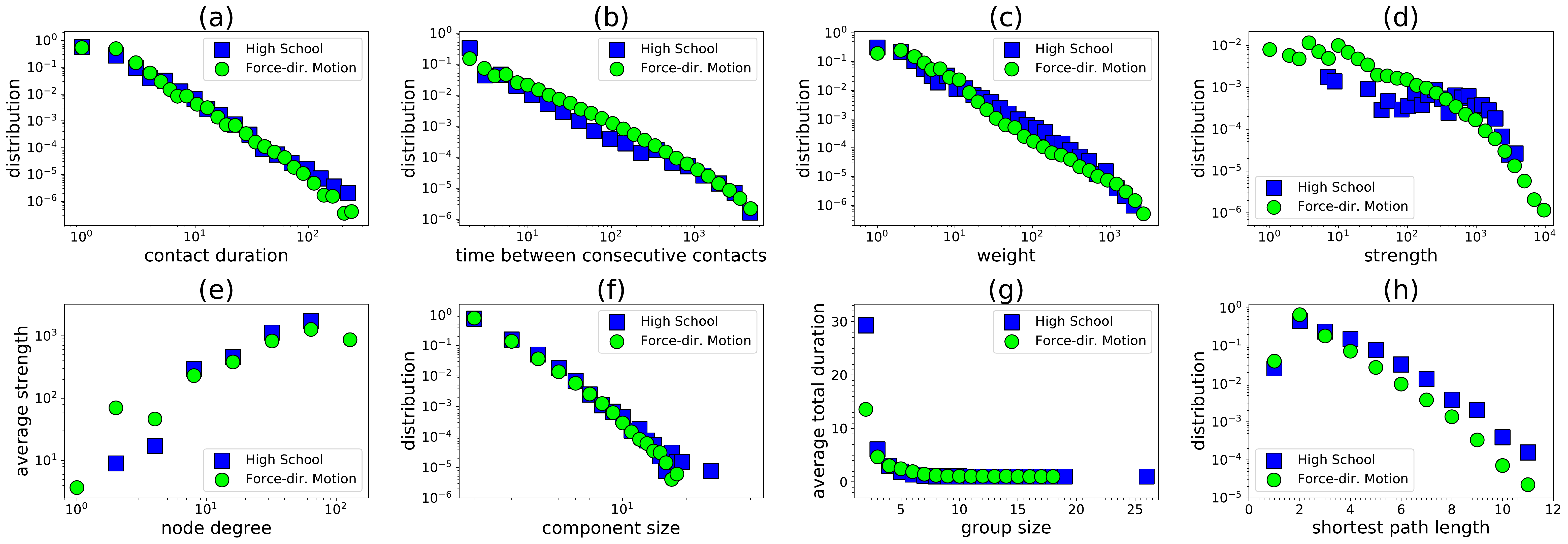}
\caption{Same as Fig.~\ref{figHPProps_Popularity} but for the High School.
\label{figHSProps_Popularity}}
\end{figure}

\begin{figure}[!h]
\centering
\includegraphics[width=17.8cm,height=6.2cm]{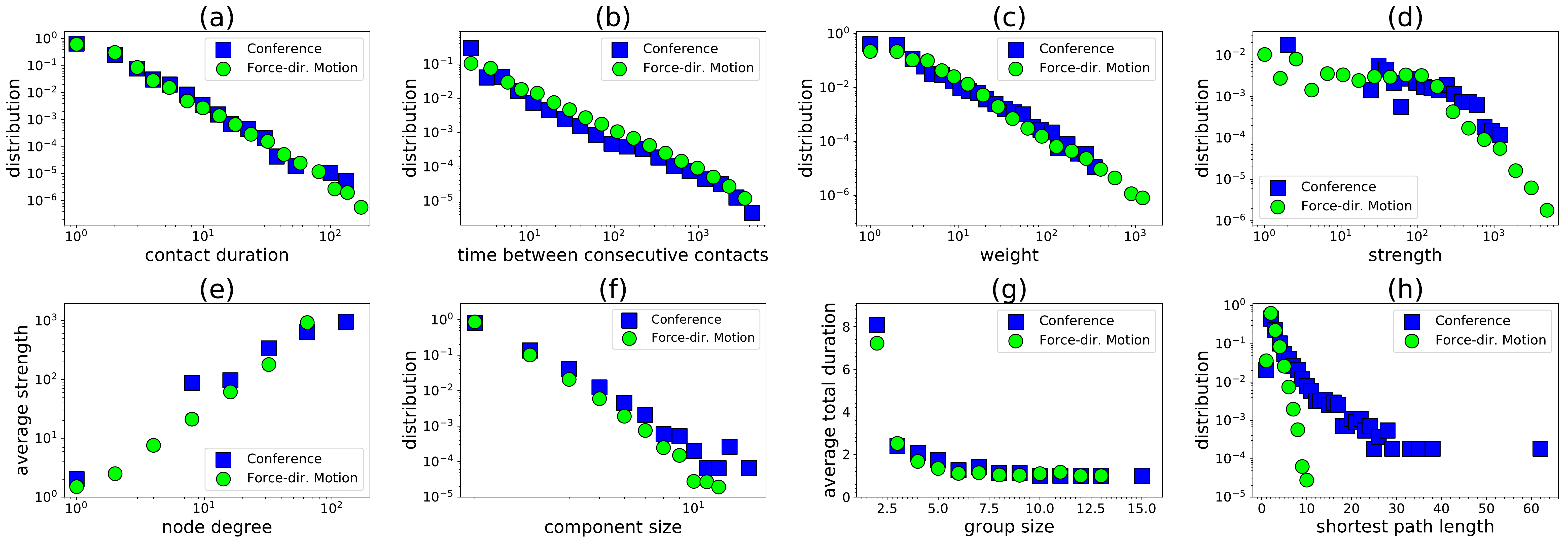}
\caption{Same as Fig.~\ref{figHPProps_Popularity} but for the Conference.
\label{figHYTProps_Popularity}}
\end{figure}

One can replace the angular distances $s_{ij}=R\Delta\theta_{ij}$ with effective distances $\chi_{ij}=s_{ij}/(\kappa_{i} \kappa_{j})$ in the bonding and attractive forces of the FDM (Eqs.~(\ref{eq:escape_prob}), (\ref{eq:ForceEq}) in the main text). However, in all datasets we considered the distribution of $\kappa$s was in general quite homogeneous to justify the need for this description---see Fig.~\ref{figKappas}, where the expected degree $\kappa$ of each agent is its average degree per time slot. Indeed, in Figs.~\ref{figHPComps_Popularity}-\ref{figHYTProps_Popularity} we see that if we assign to agents the estimated $\kappa$s from the real data and use effective distances in the FDM, we obtain very similar results as in Secs.~\ref{sec:recurrent_components},~\ref{SecProps} where we use only angular distances. For the simulations in Figs.~\ref{figHPComps_Popularity}-\ref{figHYTProps_Popularity} we tune again the model parameters $L$, $\mu_1$, $F_0$, $\mu_2$ as described in Sec.~\ref{sec:parameter_tuning}, see Table \ref{tableEffectiveParams} for their values. The rest of the simulation parameters are as in Sec.~\ref{sec:parameter_tuning}.

\begin{table}[!h]
\begin{tabular}{|c|c|c|c|c|c|}
\hline 
Network & $T_{\textnormal{warmup}}$ & $L$ & $\mu_1$ & $F_0$ & $\mu_2$  \\ 
\hline 
Hospital & 2500 & 95 & 29 & 0.12 & 33  \\ 
\hline 
Primary School & 2000 & 105 & 2.7 & 0.2 & 6.1  \\ 
\hline 
High School & 6500 & 240 & 23 & 0.2 & 16  \\ 
\hline 
Conference & 1800 & 75 & 145 & 0.05 & 85  \\ 
\hline 
\end{tabular}
\caption{Parameter values in the simulations with the FDM model that uses effective distances.
\label{tableEffectiveParams}}
\end{table}


%

\end{document}